\documentclass[12pt]{article}
\usepackage{epsfig,amsmath,amsfonts,amssymb,amstext,afterpage,psfrag,
}
\usepackage{slashed}
\usepackage[square,comma,sort&compress,numbers]{natbib}
\usepackage{overpic,subfigure,yfonts}
\allowdisplaybreaks
\def\subhat{\raisebox{-18pt}{$\hat{}$}}

\def\spose#1{\hbox to 0pt{#1\hss}}
\def\lsim{\mathrel{\spose{\lower 3pt\hbox{$\mathchar"218$}}
 \raise 2.0pt\hbox{$\mathchar"13C$}}}
\def\gsim{\mathrel{\spose{\lower 3pt\hbox{$\mathchar"218$}}
 \raise 2.0pt\hbox{$\mathchar"13E$}}}

\begin{document}

\begin{titlepage}

\begin{flushright}
{\small
LMU-ASC~35/16\\ 
August 2016
}
\end{flushright}

\vspace{0.5cm}
\begin{center}
{\Large\bf \boldmath                                               
Standard Model Extended by a Heavy Singlet:\\
\vspace*{0.3cm}
Linear vs. Nonlinear EFT  
\unboldmath}
\end{center}

\vspace{0.5cm}
\begin{center}
{\sc G.~Buchalla, O.~Cat\`a, A.~Celis and C.~Krause} 
\end{center}

\vspace*{0.4cm}

\begin{center}
Ludwig-Maximilians-Universit\"at M\"unchen, Fakult\"at f\"ur Physik,\\
Arnold Sommerfeld Center for Theoretical Physics, 
D--80333 M\"unchen, Germany
\end{center}

\vspace{1.5cm}
\begin{abstract}
\vspace{0.2cm}
\noindent
We consider the Standard Model extended by a heavy scalar singlet in 
different regions of parameter space and construct the appropriate 
low-energy effective field theories up to first nontrivial order. 
This top-down exercise in effective field theory is meant primarily to 
illustrate with a simple example the systematics of the linear and nonlinear 
electroweak effective Lagrangians and to clarify the relation between them. 
We discuss power-counting aspects and the transition between both effective 
theories on the basis of the model, confirming in all cases the rules 
and procedures derived in previous works from a bottom-up approach.
\end{abstract}

\vfill

\end{titlepage}

\section{Introduction}
\label{sec:intro}

The discovery of the Higgs boson at the LHC together with the absence (so far) 
of new-physics states has triggered a renewed interest in effective field 
theories (EFTs) at the electroweak scale. In the last years, there has been a 
surge of papers reassessing different technical and conceptual aspects 
(completeness of operators~\cite{Grzadkowski:2010es,Buchalla:2013rka}, 
aspects of power counting~\cite{Buchalla:2013eza,Buchalla:2016sop}, etc.), 
and a program to carry out the one-loop renormalization of the EFTs has 
emerged~\cite{Jenkins:2013zja,Jenkins:2013wua,Alonso:2013hga,Guo:2015isa}. 
This has been paralleled by an increasing interest in exploiting the potential 
of EFTs as a phenomenological tool for indirect searches of new physics at 
the LHC~\cite{Passarino:2012cb,Carmi:2012yp,Azatov:2012bz,Buchalla:2015wfa,Buchalla:2015qju}. 
One of the main goals of the recent 
developments is to get the formalism ready for the level of scrutiny required 
at the LHC in the forthcoming Run II and III (see, e.g.,~\cite{YellowReport} 
for an updated review).  

The main virtue of an EFT approach is that it is general and 
model-independent. Once (i) the symmetries and the particle content relevant 
at the scale of interest and (ii) the nature of the underlying dynamics are 
specified, the resulting set of operators represents the most general way in 
which deviations caused by ultraviolet (UV) physics can be parametrized. 
If the UV physics is known, one can construct the EFT by integrating 
out the heavy degrees of freedom. This is sometimes referred to as a 
{\emph{top-down}} approach. EFTs of this sort are typically useful to simplify 
calculations at low scales. More challenging are those situations where the 
ultraviolet physics is unknown. Such {\emph{bottom-up}} EFTs heavily rely on 
(i) symmetry arguments for the build-up of operators and (ii) power counting 
both in order to organize the expansion and to estimate the typical size of 
the operator coefficients. By comparing the estimated sizes of operators with 
their experimental bounds one is thus sensitive to indirect effects from 
new physics.   
 
In the electroweak sector, there are two different (bottom-up) EFTs one can 
build. They both are invariant under the Standard Model gauge symmetry and 
have the same particle content. However, they fundamentally differ in the 
assumed nature of the dynamics responsible for electroweak symmetry breaking. 
As a result, the very nature of the EFT expansion, i.e. its power counting, 
is different. If the underlying dynamics is weakly coupled, new-physics 
effects decouple and the expansion is in canonical dimensions of the fields. 
In contrast, if the underlying dynamics is strongly coupled (around the TeV 
scale), new-physics effects do not decouple and the expansion is topological 
(i.e., in the number of loops), or equivalently in the chiral dimensions of 
fields {\emph{and}} couplings~\cite{Buchalla:2013eza}. 

These two EFTs are normally termed linear and nonlinear, in reference to the 
realization of the electroweak gauge symmetry. In the former, the scalar 
sector is most conveniently assembled as an electroweak doublet field 
$\Phi(x)$, while in the latter it is convenient to split the 
Goldstone modes and the Higgs scalar and represent them with the fields $U(x)$ 
and $h(x)$, respectively. Obviously the choice of variables is a matter of 
convention: physics certainly should not depend on how the scalar degrees of 
freedom are parametrized. The choice of variables simply makes the power 
counting associated with each EFT more transparent.  

In this paper we would like to show this difference in power counting 
explicitly from a top-down approach, using a simple UV-complete toy model and 
integrating out its heavy degrees of freedom. This model should be rich enough 
to possess, depending on the values of its parameters, a decoupling and 
nondecoupling regime while still being perturbative. We examine the 
simplest model that exhibits these features, namely the Standard Model 
extended with a heavy real scalar field endowed with a $Z_2$ 
symmetry~\cite{Schabinger:2005ei,Barbieri:2007bh,Barger:2007im,Pruna:2013bma,Robens:2015gla,Falkowski:2015iwa,Buttazzo:2015bka,Cheung:2015dta,Corbett:2015lfa,Bojarski:2015kra,Ghosh:2015apa,Robens:2016xkb,Boggia:2016asg,Feruglio:2016zvt}. 
If the heavy field acquires 
a nontrivial vacuum expectation value, this model can be recast as a $SO(5)$ 
linear sigma model both spontaneously and explicitly broken down to $SO(4)$. 
We show explicitly how, depending on the sizes of the different 
parameters, integrating out the heavy scalar generates either a nonlinear EFT 
(with a pseudo-Goldstone Higgs) or a linear EFT (with a Standard Model Higgs), 
leading to expansions in either chiral or canonical dimensions. 

From a phenomenological viewpoint, this scalar model is far from being 
realistic as an extension of the Standard Model. On the one hand, current 
experimental Higgs data severely constrain its parameter 
space~\cite{Robens:2015gla,Robens:2016xkb}, especially in the nondecoupling 
regime. On the other hand, a realistic strongly-coupled sector is likely to 
be more sophisticated, with a confining phase giving rise to an infinite set 
of resonances, much like what happens in QCD. However, even in QCD the 
(linear) sigma model, while not phenomenologically realistic, is still 
useful to the extent that it illustrates the systematics of the corresponding 
low-energy expansion, chiral perturbation theory (ChPT). 
In this paper, we follow a similar strategy for the electroweak sector. 
The value of the toy model is therefore not its phenomenological viability, 
but the fact that it illustrates in a simple and explicit way how the 
linear and nonlinear EFTs are related.  

Interestingly, the scalar toy model not only clarifies the origin of the 
different power countings, but also shows that in certain settings the 
transition between a nonlinear and a linear EFT is not a discrete choice but 
a continuous one. In particular, there is a well-defined limit, in which the 
Standard Model is recovered. This supports the claim
\cite{Buchalla:2015wfa,Buchalla:2015qju} that using a nonlinear EFT at the 
LHC is the right framework to determine the nature of the Higgs boson from 
experimental data. 

This paper is organized as follows: 
In Sections~\ref{sec:model} and \ref{sec:lphys} we describe the toy 
model and work out its couplings in the nonlinear Higgs representation. 
In Section~\ref{sec:nonlin}
we integrate out the heavy scalar in the nondecoupling regime. 
We work out the effective Lagrangian at tree level up to 
next-to-leading order (NLO) and find a 
particular version of the electroweak chiral Lagrangian (EWChL). 
In Section~\ref{sec:lin} we 
repeat the same steps in the weakly-coupled regime and end up with the Standard 
Model extended by dimension-6 operators. We also examine the transition
between the two different regimes. Section~\ref{sec:xi2} is devoted to the 
decoupling limit of the general, model-independent chiral Lagrangian. 
Expanding this nonlinear EFT for small values of 
$\xi=v^2/f^2$, the ratio of scalar vacuum expectation values, to 
${\cal{O}}(\xi^n)$, one recovers the expansion of the linear EFT to 
operators of dimension $d=2n+4$. We do this explicitly for the 
leading-order (LO) chiral Lagrangian through ${\cal{O}}(\xi^2)$. 
We summarize our conclusions in Section~\ref{sec:concl}. 
Technical details are relegated to the Appendix.

\section{Model}
\label{sec:model}

We consider an extension of the Standard Model (SM)
with the Higgs doublet $\Phi$ by a real scalar gauge singlet $S$.
Imposing a $Z_{2}$ symmetry under which $S \rightarrow - S$, 
the Lagrangian for the scalar sector reads~\cite{Schabinger:2005ei,Barbieri:2007bh,Barger:2007im,Pruna:2013bma,Robens:2015gla,Falkowski:2015iwa,Buttazzo:2015bka,Cheung:2015dta,Corbett:2015lfa,Bojarski:2015kra,Ghosh:2015apa,Robens:2016xkb,Boggia:2016asg,Feruglio:2016zvt}
\begin{equation}
\label{eq:2.1}
\mathcal{L} = (D^{\mu} \Phi)^{\dag} (D_{\mu} \Phi)  +  
\partial^{\mu}  S \partial_{\mu} S   - V(\Phi,S) 
\end{equation}
with
\begin{equation}
\label{eq:2.2}
V(\Phi,S) =  -\frac{ \mu_1^2}{2} \Phi^{\dag} \Phi  - \frac{\mu_2^2}{2} S^2  + 
\frac{\lambda_1}{4}  ( \Phi^{\dag} \Phi )^2  + \frac{\lambda_2}{4}  S^4 + 
\frac{\lambda_3}{2}    \Phi^{\dag} \Phi  S^2 
\end{equation}
Requiring the potential to be bounded from below
and to have a stable minimum implies
\begin{equation}
\label{eq:2.8}
  \lambda_1,  \,  \lambda_2 >0\,, \qquad   
\lambda_1  \lambda_2 - \lambda_3^2 > 0  
\end{equation}
The scalar fields develop vacuum expectation values (vevs), 
\begin{equation}\label{phih1u}
\Phi = \frac{v+h_1}{\sqrt{2}} U 
\left(\begin{array}{c} 0\\ 1\end{array}\right)\, ,\qquad
S = \frac{v_{s} + h_2}{\sqrt{2}}
\end{equation}
Here we write $\Phi$ in polar coordinates, where $U=\exp(2i\varphi^a T^a/v)$ 
is the Goldstone-boson matrix. The vevs are given by
\begin{align}
\label{eq:2.3}
\mu_1^2 =& \frac{   \lambda_1 v^2 +\lambda_3 v_{s}^2  }{  2  } \,, \qquad 
\mu_2^2 = \frac{   \lambda_3 v^2 +\lambda_2 v_{s}^2  }{  2  } 
\end{align}
We obtain the physical states after the rotation
\begin{equation}\label{eq:2.4}
\left(\begin{array}{c} h\\ H\end{array}\right)\; = \;
\left[\begin{array}{cc} \cos{\chi} & -\sin{\chi} \\ 
    \sin{\chi} & \cos{\chi}\end{array}\right]\;
\left(\begin{array}{c} h_1\\ h_2\end{array}\right)  
\end{equation}
with
\begin{equation}
\label{eq:2.5}
\tan(2 \chi) = \frac{2\lambda_3   v  v_{s} }{\lambda_2  v_{s}^2 -\lambda_1 v^2 }
\end{equation}
Without loss of generality we may restrict the range of $\chi$ to 
$-\pi/2 \leq \chi \leq \pi/2 $. The masses of the scalar bosons are
\begin{equation}
\label{eq:2.6}
M_{h, H}^2 = \frac{1}{4} \left[\lambda_1 v^2 +    \lambda_2  v_{s}^2 \mp 
\sqrt{(\lambda_1 v^2 -\lambda_2  v_{s}^2)^2 + 4(\lambda_3  v  v_{s}  )^2}\right]
\end{equation}
with $M_h\equiv m< M_H\equiv M$ by convention.     

The full parameter space of the model in (\ref{eq:2.1}) is spanned by the
five values of $\mu_1$, $\mu_2$, $\lambda_1$, $\lambda_2$ and $\lambda_3$.
Equivalently, we may express those in terms of the physical quantities 
$m$, $v$, $M$, $f\equiv\sqrt{v^2+v^2_s}$ and $\chi$, or
\begin{equation}\label{mvrxiom}
m,\quad v,\quad r\equiv\frac{m^2}{M^2},\quad \xi\equiv\frac{v^2}{f^2},
\quad \omega\equiv\sin^2\chi
\end{equation}
The two sets of parameters are related through
\footnote{Note that $\sin\chi\equiv{\rm sgn}(\chi)\sqrt{\omega}$. In the 
following, we sometimes write $\sin\chi=\sqrt{\omega}$ for simplicity,
dropping the ${\rm sgn}(\chi)$, which has to be included for negative $\chi$.}
\begin{align}
\label{eq:2.7}
\lambda_1 =&  \frac{2 M^2}{ f^2} \frac{  r + \omega (1-r)  }{  \xi } 
\nonumber \\ 
\lambda_2 =& \frac{2 M^2}{ f^2} \frac{   1 - \omega (1-r) }{   1 - \xi } 
\nonumber \\
\lambda_3 =& \frac{ 2 M^2 }{f^2}   (1-r)  
\sqrt{ \frac{   \omega (1- \omega) }{  \xi (1- \xi) }    }
\end{align}
together with (\ref{eq:2.3}).
After fixing $v=(\sqrt{2} G_F)^{-1/2}=246$~GeV and $m=125$~GeV in 
(\ref{mvrxiom}), we are left with $r$, $\xi$ and $\omega$, parametrizing
the dynamics beyond the SM. 
Apart from the resonance mass $M$, which sets the scale of new-particle
thresholds, and which we assume to be in the TeV range, this dynamics is 
essentially governed by the two parameters $\xi$ and $\omega$, 
where $\xi,\, \omega \in [0,1]$.

\vspace*{0.5cm}

Unless specified otherwise, we typically assume a situation where the scalar 
sector exhibits an approximate $SO(5)$ symmetry.
Under this symmetry the four real components of $\Phi$ and $S$
transform in the fundamental representation. This limit is
physically motivated as the Higgs mass $m$ is then protected by
the pseudo-Goldstone nature of the field $h$, which is of interest in
particular in the strongly-coupled scenario  \cite{Agashe:2004rs}.

In the strict $SO(5)$ symmetric limit, we 
have $\lambda_1 = \lambda_2 = \lambda_3 \equiv \lambda= 2 M^2/f^2$, 
$r=0$ and $\omega = \xi$. Also in this limit $\mu_1 =\mu_2 = M$. 
We parametrize deviations from the exact symmetry by 
$r$ and $\delta \equiv \omega/\xi -1$.
We denote by
\begin{equation}\label{sigdef}
\Sigma^2 \equiv \Phi^\dagger\Phi + S^2
\end{equation}
the square of the scalar multiplet in the fundamental representation 
of $SO(5)$.
We then decompose the potential (\ref{eq:2.2}) as $V\equiv V_0+V_1$ into
an $SO(5)$ invariant part,
\begin{equation}\label{v0def}
V_0=-\frac{\mu^2_1}{2} \Sigma^2 +\frac{\lambda_1}{4}\Sigma^4
\end{equation}
and terms that explicitly break the $SO(5)$ symmetry,
\begin{equation}\label{v1def}
V_1=\frac{\mu^2_1-\mu^2_2}{2} S^2 +\frac{\lambda_1+\lambda_2-2\lambda_3}{4} S^4
+\frac{\lambda_3-\lambda_1}{2} \Sigma^2 S^2
\end{equation}
The three $SO(5)$-breaking couplings in (\ref{v1def})
correspond to the three different $SO(5)$-breaking, $SO(4)$-symmetric 
operators of dimension less or equal to four that
respect the $Z_2$ symmetry of the model: $S^2$, $S^4$, and $\Sigma^2 S^2$.
All three are governed by the $SO(5)$-breaking operator $S\equiv n^T\Sigma$,
where $n^T=(0,0,0,0,1)$ is the spurion that breaks $SO(5)$ while preserving
$SO(4)$.

For small $SO(5)$ breaking, the case of particular interest to us,
we require $r$, $\delta\ll 1$. 
Expanding the couplings in (\ref{v1def}) to first order in $r$ and $\delta$,
we find, using (\ref{eq:2.3}) and (\ref{eq:2.7}),
\begin{align}\label{v1coup}
\mu^2_1-\mu^2_2 =& M^2 \frac{\delta}{2(1-\xi)} \nonumber \\
\lambda_1+\lambda_2 -2\lambda_3 =& \frac{2 M^2}{f^2}\frac{r}{\xi(1-\xi)}
\nonumber\\
\lambda_3-\lambda_1 =& -\frac{2M^2}{f^2}\left(\frac{r}{\xi}+
\frac{\delta}{2(1-\xi)}\right)
\end{align}
The requirement $r$, $\delta\ll 1$ ensures that the dimensionless
couplings in (\ref{v1def}) remain weak (of order unity) even for
large $\lambda_i$. Similarly, $\mu^2_1-\mu^2_2$ remains of order
$v^2$ for large $M^2$.

Counting parameters, we observe that we can group the five couplings of 
the original potential (\ref{eq:2.2}) into the two $SO(5)$-symmetric
couplings in (\ref{v0def}) and the three $SO(5)$-breaking couplings
in (\ref{v1def}). The former correspond to $M$ and $f$, the latter to
$r$, $\delta$ and $\xi$. Out of these three, $r$ and $\delta$ control
the (small) $SO(5)$ breaking, whereas $\xi$ is naturally of order unity.
The last property reflects the degeneracy of vacua in the strict
$SO(5)$ limit, which is lifted by the small explicit symmetry breaking
triggered by $r$ and $\delta$.

\vspace*{0.5cm}

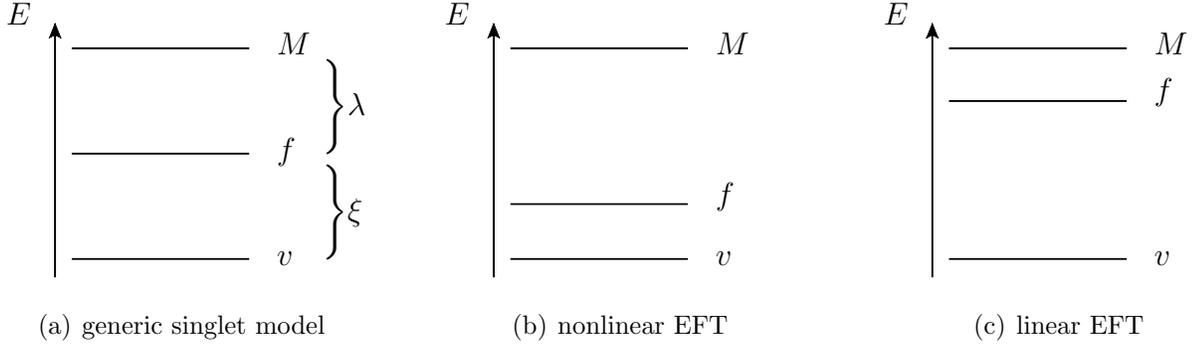
\begin{figure*}[t]
\begin{center}
\subfigure[generic singlet model\label{fig:scalesC}]{
\begin{overpic}[width=4cm]{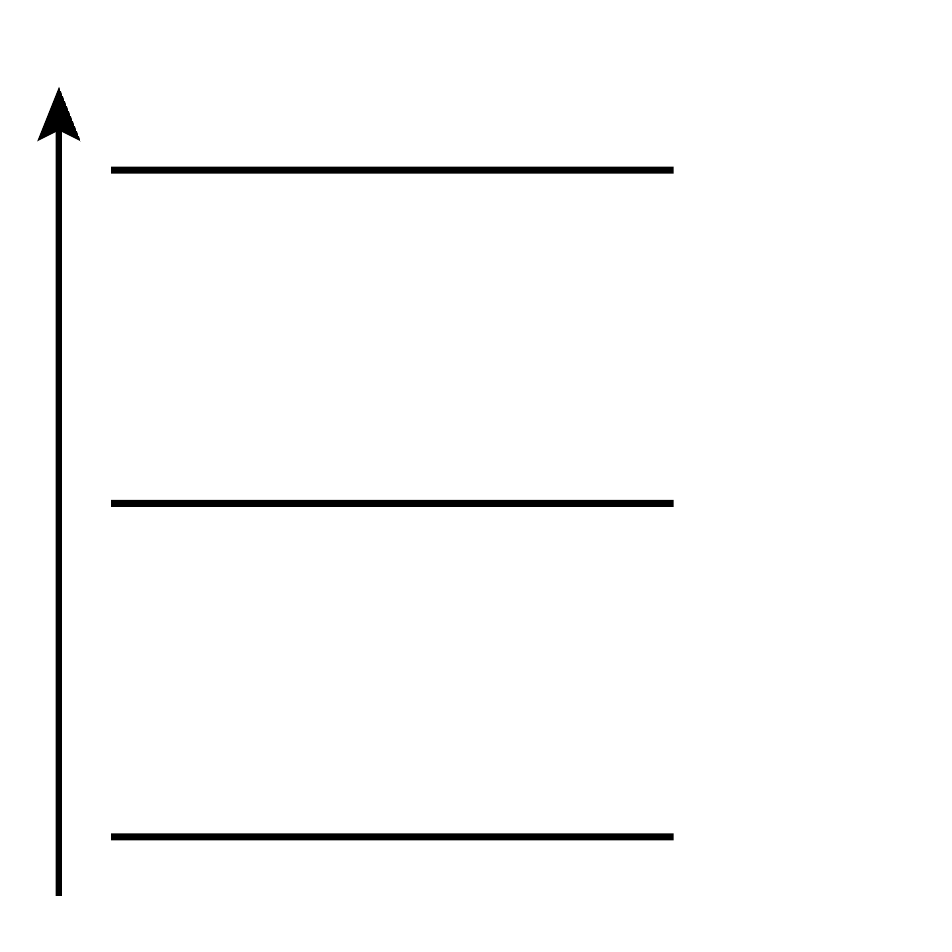}
\put (-10,90){$E$}
\put (80,80){$M$}
\put (95,60){$\Bigg\}\lambda$}
\put (80,45){$f$}
\put (95,25){$\Bigg\}\xi$}
\put (80,10){$v$}
\end{overpic}}\hfill
\subfigure[nonlinear EFT\label{fig:scalesA}]{
\begin{overpic}[width=4cm]{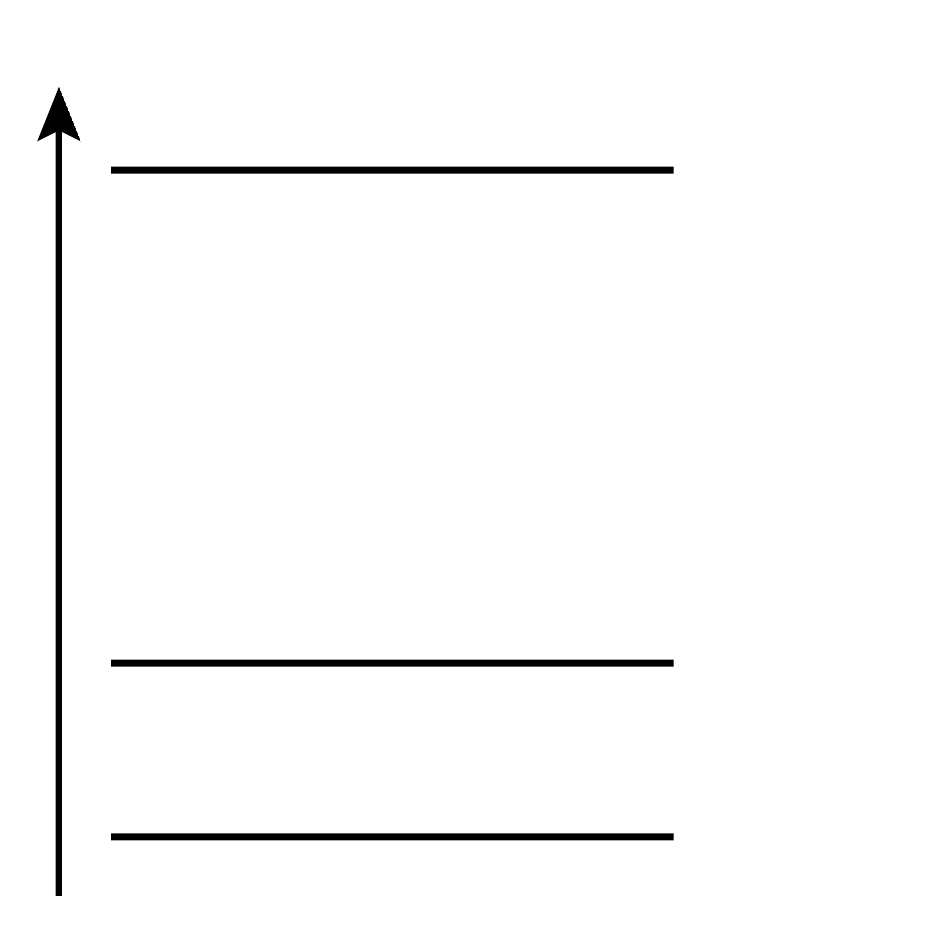}
\put (-10,90){$E$}
\put (80,80){$M$}
\put (80,30){$f$}
\put (80,10){$v$}
\end{overpic}}\hfill
\subfigure[linear EFT\label{fig:scalesB}]{
\begin{overpic}[width=4cm]{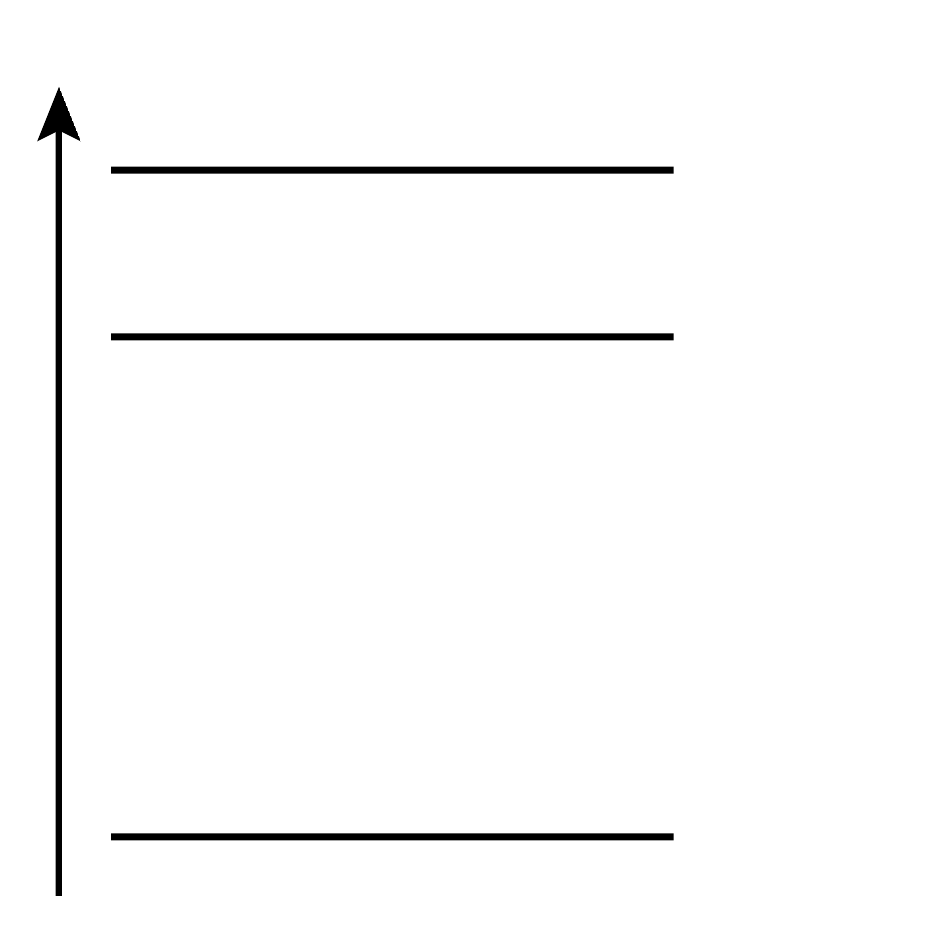}
\put (-10,90){$E$}
\put (80,80){$M$}
\put (80,65){$f$}
\put (80,10){$v$}
\end{overpic}}
\end{center}
\caption{Schematic picture of the different possible hierarchies. 
Further details are given in the main text.}
\label{fig:scales}
\end{figure*}

For the construction of a low-energy EFT by 
integrating out high mass scales, we are mainly interested
in the following two basic scenarios, 
depicted in Figs. \ref{fig:scalesA} and \ref{fig:scalesB}:
\begin{description}
\item I) strongly-coupled regime (nonlinear EFT)
\begin{equation}\label{slimit}
|\lambda_i| \lsim 32\pi^2,\quad  m\sim v \sim f \ll M
\qquad\Rightarrow\qquad  \xi, \omega ={\cal O}(1)
\end{equation}
\item II) weakly-coupled regime (linear EFT)
\begin{equation}\label{wlimit}
\lambda_i={\cal O}(1),\quad    m\sim v \ll f\sim M
\qquad\Rightarrow\qquad \xi,\omega \ll 1
\end{equation}
\end{description}
The nominal strong-coupling limit has $M\approx 4\pi f$, corresponding 
to $|\lambda_i|\approx 32\pi^2$. In this case, a simple description of the 
dynamics in terms of a resonance $H$ would cease to be valid. 
We assume that the $\lambda_i$ remain somewhat below, in a regime 
where perturbation theory is still a sufficiently reliable approximation.

We will show that integrating out $M$ in case I) leads to a nonlinear
EFT, organized by a power counting in chiral dimensions. We will also 
demonstrate that integrating out $M\sim f$ in case II) gives rise
to a linear EFT, organized in terms of canonical dimensions.

\section{Full scalar Lagrangian in terms of the 
physical fields}
\label{sec:lphys}

Following the notation of \cite{Buchalla:2013rka}, we write the complete
Lagrangian of the SM extended by a scalar singlet as
\begin{equation}\label{lfull}
{\cal L} = {\cal L}_0 + {\cal L}_{hH}
\end{equation}
where
\begin{equation}\label{lsm0}
{\cal L}_0 = -\frac{1}{2} \langle G_{\mu\nu}G^{\mu\nu}\rangle
-\frac{1}{2}\langle W_{\mu\nu}W^{\mu\nu}\rangle 
-\frac{1}{4} B_{\mu\nu}B^{\mu\nu}
+\bar q i\!\not\!\! Dq +\bar\ell i\!\not\!\! D\ell
+\bar u i\!\not\!\! Du +\bar d i\!\not\!\! Dd +\bar e i\!\not\!\! De 
\end{equation}
and the scalar sector is given, in terms of the physical fields $h$ and $H$, 
by
\begin{align}
\label{eq:3.1}
\mathcal{L}_{hH} &= \frac{1}{2} \partial_{\mu} h \partial^{\mu} h + 
\frac{1}{2} \partial_{\mu} H \partial^{\mu} H - V(h,H) \nonumber \\
&+ \frac{v^2}{4} \langle D_{\mu} U^{\dag} D^{\mu} U \rangle \left( 1 +  
\frac{2 c}{v}  h +  \frac{2 s }{v}  H +  \frac{c^2}{v^2}  h^2  +  
\frac{s^2}{v^2}  H^2 +  \frac{2 s c}{v^2}  h H      \right) \nonumber \\
&-v \left(\bar q Y_u U P_+ \textfrak{r} + \bar q Y_d U P_- \textfrak{r}
+ \bar \ell Y_e U P_- \eta 
+ \mathrm{h.c.}  \right)  \left[1 + \frac{c}{v} h + \frac{s}{v} H  \right]
\end{align}
Here $U=\exp(2i\varphi^a T^a/v)$ is the Goldstone-boson matrix;
$q=(u_L,d_L)^T$ and $\ell=(\nu_L,e_L)^T$ are the left-handed doublets;
$u=u_R$, $d=d_R$ and $e=e_R$ the right-handed singlets; and 
$\textfrak{r}=(u_R,d_R)^T$, $\eta=(\nu_R,e_R)^T$. 
We suppress generation indices. The coefficients are
\begin{equation}
\label{eq:3.2}
\cos \chi\equiv c \,, \qquad \sin \chi\equiv s 
\end{equation}
The full scalar potential reads
\begin{align}
\label{eq:3.3}
V(h,H) =&  \frac{1}{2} m^2 h^2 +   \frac{1}{2} M^2 H^2  - d_1 h^{3} - 
d_2  h^2 H - d_3 h H^2  - d_4 H^3 \nonumber \\
&- z_1 h^4 - z_2 h^3 H - z_3 h^2 H^2 -z_4 h H^3 - z_5 H^4  
\end{align}
with

\begin{align}
d_1&=\frac{m^2}{2vv_s}[s^3 v-c^3 v_s]\nonumber\\
d_2&=-\frac{2m^2+M^2}{2vv_s}s c [s v+ c v_s]\nonumber\\
d_3&=\frac{2M^2+m^2}{2vv_s}s c[c v-s v_s]\nonumber\\
d_4&=-\frac{M^2}{2vv_s}[c^3 v+ s^3 v_s]\nonumber\\
z_1&=-\frac{1}{8v^2v_s^2}\left[m^2(s^3 v-c^3 v_s)^2+
 M^2 s^2 c^2 (s v+c v_s)^2\right]\nonumber\\
z_2&=\frac{s c}{2v^2v_s^2}(s v+c v_s)\left[m^2(s^3 v- c^3 v_s)+
 M^2 s c (c v-s v_s)\right]\nonumber\\
z_3&=-\frac{s c}{8v^2v_s^2}\left[m^2(6 s c (s v+c v_s)^2-2vv_s)+
 M^2(6 s c(c v-s v_s)^2+2vv_s)\right]\nonumber\\
z_4&=\frac{s c}{2v^2v_s^2}(c v -s v_s)\left[M^2(c^3 v+s^3 v_s)+
 m^2 s c (s v+c v_s)\right]\nonumber\\
z_5&=-\frac{1}{8v^2v_s^2}\left[M^2(c^3 v+ s^3 v_s)^2+
 m^2 s^2 c^2 (c v - s v_s)^2\right]
\end{align} 

We emphasize that (\ref{eq:3.1}) represents the complete, renormalizable 
model, expressed here in terms of nonlinear coordinates $U$ for the 
electroweak Goldstone fields. 

\section{Nonlinear EFT limit}
\label{sec:nonlin}

In this section we integrate out the heavy scalar mass eigenstate $H$ 
at tree level in the strongly-coupled limit defined in (\ref{slimit}),
including leading and next-to-leading order terms.
We show that the resulting EFT takes the form of the electroweak chiral
Lagrangian with a light Higgs
\cite{Buchalla:2013rka,Buchalla:2013eza,Feruglio:1992wf}.
To leading order the scalar sector of this Lagrangian can, in general, 
be written as \cite{Buchalla:2013rka,Buchalla:2013eza} 
\begin{align}  \label{eq:LO}
\mathcal{L}_{Uh,LO} =& \frac{v^2}{4} \langle D_{\mu} U^{\dag} D^{\mu} U \rangle 
\left( 1 + F_U(h)\right) + \frac{1}{2} \partial_{\mu} h \partial^{\mu} h - V(h) 
\nonumber \\ 
&- v \Bigl[Ê \bar q \left(  Y_u + \sum_{n=1}^{\infty}  Y_{u}^{(n)} 
\left( \frac{h}{v} \right)^n  \right) U P_+ \textfrak{r}   + \bar q    
\left( Y_d + \sum_{n=1}^{\infty} Y_{d}^{(n)} 
\left(  \frac{h}{v} \right)^n  \right) U P_- \textfrak{r}   \nonumber \\
& + \bar \ell \left( Y_e + \sum_{n=1}^{\infty} Y_e^{(n)} 
\left( \frac{h}{v}  \right)^n \right) U P_- \eta + \mathrm{h.c.} \Bigr]
\end{align}
to be supplemented by the usual gauge and fermion terms of the unbroken 
SM (\ref{lsm0}). 

We start from the full theory in (\ref{lfull}) and follow the procedure
outlined in \cite{Buchalla:2013rka}.
The part of this Lagrangian that depends on $H$ reads
\begin{equation}\label{eq:4.1}
{\cal L}_H =\frac{1}{2}H(-\partial^2-M^2)H
+J_1 H + J_2 H^2 + J_3 H^3 + J_4 H^4
\end{equation}
where the $J_i$ are given by
\begin{align}\label{jji}
J_1 =& d_2 h^2 + z_2 h^3 +\frac{v^2}{4}\langle D_{\mu} U^{\dag} D^{\mu} U \rangle 
\left(\frac{2 s}{v} +\frac{2 s c}{v^2}h\right) - s J_f\nonumber\\
J_2 =& d_3 h + z_3 h^2 +\frac{s^2}{4}\langle D_{\mu} U^{\dag} D^{\mu} U \rangle 
\nonumber\\
J_3 =& d_4 + z_4 h\, ,\qquad\qquad J_4=z_5
\end{align}
with
\begin{equation}\label{jfdef}
J_f\equiv \bar q Y_u U P_+ \textfrak{r} + \bar q Y_d U P_- \textfrak{r} + 
\bar \ell Y_e U P_- \eta + \mathrm{h.c.} 
\end{equation}
To perform the EFT expansion, 
we make the dependence of the $J_i$ on the heavy mass $M$
explicit by writing
\begin{equation}\label{eq:4.2}
J_i\equiv M^2 J^0_i +\bar J_i
\end{equation}
and similarly
\begin{equation}\label{dizi}
d_i\equiv M^2 d_{i0} +\bar d_i\, ,\qquad
z_i\equiv M^2 z_{i0} +\bar z_i
\end{equation}
for the coefficients in the potential (\ref{eq:3.3}).
The $J^0_i$ are pure polynomials in $h$.

We integrate out the heavy field $H$ at tree level by solving 
its equation of motion 
\begin{equation}\label{eq:4.3}
(-\partial^2-M^2 + 2 J_2) H +J_1 + 3 J_3 H^2 + 4 J_4 H^3 = 0
\end{equation}
and inserting the solution into the Lagrangian (\ref{eq:3.1}). 
We can solve (\ref{eq:4.3}) order by order in powers of $1/M^2$ 
by expanding
\begin{equation}\label{eq:4.4}
H = H_0 + H_1 + H_2 + \ldots,\qquad\quad 
H_l = {\cal O}(1/M^{2l})
\end{equation}

Inserting (\ref{eq:4.4}) into (\ref{eq:4.3}) and keeping only the terms
of ${\cal O}(M^2)$ yields an (algebraic) equation for $H_0$:
\begin{equation}\label{eh0m2}
J^0_1 +(-1+2 J^0_2) H_0 + 3 J^0_3 H^2_0 + 4 J^0_4 H^3_0 = 0
\end{equation}
Retaining the terms of ${\cal O}(1)$ gives an equation that determines
$H_1$ as a function of $H_0$. The solution reads
\begin{equation}\label{h1sol}
H_1=\frac{(-\partial^2 +2\bar J_2)H_0+ \bar J_1+ 3 \bar J_3 H^2_0
+4\bar J_4 H^3_0}{M^2(1-2 J^0_2 - 6 J^0_3 H_0 -12 J^0_4 H^2_0)}
\end{equation}
Proceeding to higher orders in $1/M^2$, the $H_l$, $l\geq 2$, can be
successively computed.
 
As a first step, we obtain $H_0$ from (\ref{eh0m2}).
Since the coefficients $J^0_i$ depend on no other field than $h$, 
the solution $H_0$ will also have this property.   
It is convenient to find $H_0(h)$ as an infinite series in powers of $h$
\begin{equation}\label{eq:4.5}
H_0(h) =\sum^\infty_{k=2} r_k h^k
\end{equation}
Inserting (\ref{eq:4.5}) into (\ref{eh0m2}), we obtain for the
first few coefficients $r_k$
\begin{align}\label{r234}
r_2 &= d_{20} \nonumber\\
r_3 &= d_{20} d_{30} \nonumber\\
r_4 &= d_{20} d^2_{30} + d^2_{20} d_{40}\nonumber\\
r_5 &= d_{20} d^3_{30} + 3 d^2_{20} d_{30} d_{40}
\end{align}
In Appendix \ref{sec:exh0}, we derive a closed-form solution for $H_0(h)$ 
to all orders in $h$. We also show there that only one 
solution of the cubic equation (\ref{eh0m2}) is relevant. This solution 
starts at order $h^2$, as anticipated in (\ref{eq:4.5}).

To obtain the leading-order effective Lagrangian,
we insert $H=H_0+H_1$ into (\ref{eq:3.1}) and expand the expression,
retaining terms of ${\cal O}(M^2)$ and ${\cal O}(1)$.
Terms with $H_1$ vanish at this order due to the equation of motion for
$H_0$. We show in Appendix \ref{sec:exh0} that in general all terms of  
${\cal O}(M^2)$ cancel up to an irrelevant constant.
The leading-order scalar Lagrangian then becomes
\begin{align}\label{llo}
{\cal L}_{hH,LO} = &\frac{1}{2}(\partial h)^2 -\frac{m^2}{2} h^2
+ d_1 h^3 + \bar z_1 h^4 +\frac{1}{2}(\partial H_0)^2 
+ \bar J_1 H_0 +\bar J_2 H^2_0 +\bar J_3 H^3_0 +\bar J_4 H^4_0\nonumber\\
&+ \frac{v^2}{4} \langle D_{\mu} U^{\dag} D^{\mu} U \rangle \left( 1 +  
\frac{2 c}{v}  h + \frac{c^2}{v^2}  h^2  \right) 
- v J_f \left(1 + \frac{c}{v} h \right)
\end{align}
where $H_0=H_0(h)$. The kinetic term for $h$ has acquired the form
\begin{equation}\label{lhkin}
{\cal L}_{h,kin}=\frac{1}{2}(\partial h)^2 +\frac{1}{2}(\partial H_0)^2
=\frac{1}{2}(\partial h)^2 (1+F_h(h)) \qquad {\rm with}
\quad F_h(h) =\left(\frac{dH_0(h)}{dh}\right)^2
\end{equation}
The field redefinition \cite{Buchalla:2013rka}
\begin{equation}\label{hhtilde}
\tilde h=\int^h_0\sqrt{1+F_h(s)}\, ds = h\left(1+\frac{2}{3}r^2_2 h^2 +
\frac{3}{2}r_2 r_3 h^3+{\cal O}(h^4)\right)
\end{equation}
brings (\ref{lhkin}) to its canonical form 
${\cal L}_{h,kin}=(\partial\tilde h)^2/2$.

Eliminating $h$ in (\ref{llo}) in favour of $\tilde h$ using (\ref{hhtilde})
and dropping the tilde in the end, the scalar-sector Lagrangian takes the
form of (\ref{eq:LO}). Together with the gauge and fermion kinetic terms,
this is an electroweak chiral Lagrangian including a light Higgs boson.
Specifically, the general functions in (\ref{eq:LO}) are
\begin{align}\label{fuhvh}
F_U(h)&=2c\left(\frac{h}{v}\right)+\left[c^4- s^3 c\frac{v}{v_s}\right]
\left(\frac{h}{v}\right)^2-\frac{4}{3v_s^2}s^2c^3(vs + v_s c)^2
\left(\frac{h}{v}\right)^3+{\cal{O}}(h^4)\nonumber\\
V(h)&=m^2v^2\left[\frac{1}{2}\left(\frac{h}{v}\right)^2+
\frac{c^3v_s-s^3v}{2v_s}\left(\frac{h}{v}\right)^3-
\frac{19 s^2 c^2(s v+c v_s)^2-3(s^4v^2+c^4v_s^2)}{24v_s^2}
\left(\frac{h}{v}\right)^4\right.\nonumber\\
&\left.-\frac{s^2c^2(s v+c v_s)^3}{4v_s^3}
\left[3(1-2s^2)-\frac{c v_s-s v}{c v_s+s v}\right]\left(\frac{h}{v}\right)^5
+{\cal{O}}(h^6)\right]
\end{align}
and
\begin{align}\label{yukh}
Y_f+\sum_{n=1}^{\infty}Y_f^{(n)}\left(\frac{h}{v}\right)^n&=
Y_f\left[1+c \left(\frac{h}{v}\right)-s^2 c\frac{v s+v_s c}{2v_s}
\left(\frac{h}{v}\right)^2\right.\nonumber\\
&\left.-s^2 c^2 \frac{v s + v_s c}{6v_s^2}(4v s c + v_s (1-4s^2))
\left(\frac{h}{v}\right)^3+{\cal{O}}(h^4)\right]
\end{align}

To leading order in the $SO(5)$ limit ($\omega\to\xi$)
these expressions become
\begin{align}\label{fuhvhso5}
F_U(h)&=2\sqrt{1-\xi}\left(\frac{h}{v}\right)+
(1-2\xi)\left(\frac{h}{v}\right)^2-\frac{4}{3} \xi\sqrt{1-\xi}
\left(\frac{h}{v}\right)^3+{\cal{O}}(h^4)\nonumber\\
V(h)&=m^2v^2\left[\frac{1}{2}\left(\frac{h}{v}\right)^2+
\frac{1-2\xi}{2\sqrt{1-\xi}}\left(\frac{h}{v}\right)^3+
\frac{1}{1-\xi}\left( \frac{1}{8}-\frac{7}{6}\xi +\frac{7}{6}\xi^2 \right)
\left(\frac{h}{v}\right)^4\right.\nonumber\\
& \left. -\frac{\xi(1-2\xi)}{2\sqrt{1-\xi}} \left(\frac{h}{v}\right)^5
+{\cal{O}}(h^6)\right]
\end{align}
and
\begin{equation}\label{yukhso5}
Y_f+\sum_{n=1}^{\infty}Y_f^{(n)}\left(\frac{h}{v}\right)^n =
Y_f\left[1+\sqrt{1-\xi} \left(\frac{h}{v}\right) -\frac{\xi}{2}
\left(\frac{h}{v}\right)^2 -\frac{1}{6} \xi\sqrt{1-\xi} 
\left(\frac{h}{v}\right)^3+{\cal{O}}(h^4)\right]
\end{equation}

We can extend the derivation to include the NLO terms of ${\cal O}(1/M^2)$
in the effective Lagrangian
\begin{equation}\label{eq:4.7}
\mathcal{L}_{eff} =\mathcal{L}_{LO} + \Delta\mathcal{L}_{NLO}
+ \mathcal{O}\left(\frac{1}{M^4}\right), 
\qquad {\cal L}_{LO} = {\cal L}_0 + {\cal L}_{Uh,LO}
\end{equation}
Using (\ref{h1sol}), we find
\begin{equation}\label{eq:4.8}
\Delta\mathcal{L}_{NLO}=\frac{\left[(-\partial^2+2\bar J_2)H_0 + \bar J_1 +
3\bar J_3 H^2_0 + 4\bar J_4 H^3_0\right]^2}{2 M^2 (1- 2 J^0_2-6 J^0_3 H_0
-12 J^0_4 H^2_0)}
\end{equation}
The effective Lagrangian $\Delta\mathcal{L}_{NLO}$ contains operators that 
modify the leading-order Lagrangian \eqref{eq:LO} as well as a subset of the 
next-to-leading operators of \cite{Buchalla:2013rka}. 
In the notation of \cite{Buchalla:2013rka}, the NLO operators
generated by (\ref{eq:4.7}) are
\begin{equation}\label{eq:4.9}
{\cal O}_{D1}, {\cal O}_{D7}, {\cal O}_{D11};\quad
{\cal O}_{\psi S1}, {\cal O}_{\psi S2}, {\cal O}_{\psi S7},  
{\cal O}_{\psi S14}, {\cal O}_{\psi S15}, {\cal O}_{\psi S18}
\end{equation}
and their hermitean conjugates, together with
4-fermion operators coming from the square of the Yukawa bilinears
contained in $\bar J_1$. The 4-fermion operators that arise have the 
same structure as those in the heavy-Higgs model discussed 
in \cite{Buchalla:2012qq}, which are\footnote{The terms
$\mathcal{O}_{LR2}$, $\mathcal{O}_{LR4}$, $\mathcal{O}_{LR11}$ and 
$\mathcal{O}_{LR13}$ had been missed in the discussion of the
heavy-Higgs models in \cite{Buchalla:2013rka,Buchalla:2012qq}.}
\begin{eqnarray}\label{eq:4.10} 
&&\mathcal{O}_{FY1}, \mathcal{O}_{FY3}, \mathcal{O}_{FY5}, \mathcal{O}_{FY7}, 
\mathcal{O}_{FY9}, \mathcal{O}_{FY10}, \mathcal{O}_{ST5}, \mathcal{O}_{ST9}, \\
\mathcal{O}_{LR1}, \mathcal{O}_{LR2},\!\!\! &&\mathcal{O}_{LR3}, \mathcal{O}_{LR4},
\mathcal{O}_{LR8}, \mathcal{O}_{LR9}, \mathcal{O}_{LR10}, \mathcal{O}_{LR11}, 
\mathcal{O}_{LR12}, \mathcal{O}_{LR13}, \mathcal{O}_{LR17}, \mathcal{O}_{LR18}
\nonumber
\end{eqnarray}
and their hermitean conjugates, but they are now multiplied 
by functions $F_i(h/v)$.
\\

We discuss several important aspects of these results. 

\begin{itemize}
\item
The solution for $H_0(h)$ in the limit (\ref{slimit})
contains terms to all orders in $h$, with coefficients of ${\cal O}(1)$,
since $\xi$, $\omega={\cal O}(1)$ (see Appendix \ref{sec:exh0}).
Upon integrating out the heavy scalar, the function $H_0(h)$ enters the 
various terms in the effective Lagrangian. 
The singlet-model thus gives an explicit illustration of how the
all-order polynomial functions $F(h)$ are generated in the strong-coupling
limit of the underlying scalar sector. They are characteristic for the
nonlinear EFT.
\item
The leading-order Lagrangian, (\ref{eq:LO}) with (\ref{fuhvh}) and (\ref{yukh}),
is of ${\cal O}(1)$ in the $1/M$ expansion. The next-to-leading order terms
in (\ref{eq:4.8}) are of ${\cal O}(1/M^2)$.
However, the corresponding nonlinear EFT of the singlet model
is organized by {\it chiral dimensions\/}\footnote{The assignment
of chiral dimensions is $0$ for bosons, and $1$ for each derivative,
weak coupling or fermion bilinear.} \cite{Buchalla:2013eza},
rather than by canonical dimensions. This is expected on general grounds
and is further elaborated in the following items.
\item
It is easy to check that all terms of ${\cal L}_{LO}$ in (\ref{eq:LO}),
with (\ref{fuhvhso5}) and (\ref{yukhso5}),
including the gauge and fermion kinetic terms, carry chiral dimension 2. Note 
that the mass $m$ of the light Higgs counts with one unit of chiral dimension.
The smallness of $m$ can be understood as arising from an approximate
$SO(5)$ symmetry, where the small parameters of explicit $SO(5)$ breaking
act as weak couplings carrying chiral dimension.
\item
The NLO terms in (\ref{eq:4.8}) have chiral dimension 4,
consistent with the chiral counting.
Since we integrate out the heavy scalar at tree level, the contributions
shown in (\ref{eq:4.8}) have a suppression by $v^2/M^2$.
There are additional contributions to $\Delta{\cal L}_{NLO}$ from
one-loop diagrams of the full model, 
which are suppressed by a factor of $1/16\pi^2$.
In the strong-coupling limit $M\lsim 4\pi f$. Then both factors
are parametrically of comparable size: 
$v^2/M^2\gsim \xi/16\pi^2\approx 1/16\pi^2$.
\item
As mentioned above, the limit we consider here has a heavy mass
$M$ that stays somewhat below the nominal strong-coupling value
$4\pi f$. In that way, the picture of the heavy resonance as an elementary
field in the full theory is still a reasonable approximation.
A very similar limit was considered previously in the context of
integrating out a heavy SM Higgs to obtain a (Higgsless) 
electroweak chiral Lagrangian as the low-energy EFT in
\cite{Herrero:1993nc,Herrero:1994iu,Dittmaier:1995cr,Dittmaier:1995ee}.
\item
The results derived here at tree level remain stable under radiative 
corrections. We demonstrate this explicitly for the one-loop
effective potential in Appendix~\ref{sec:veff}. There, we show
that the one-loop corrections to the Higgs potential are at most
of order $M^2/(16\pi^2 f^2)$ in the case of a weakly broken
$SO(5)$ symmetry. This is smaller than one for large, but
still sufficiently perturbative couplings. In the nominal strong-coupling
case, the loop corrections would become of order unity. The potential
would then no longer be calculable, as expected.
\end{itemize}

We end this section with an illustration of how the
nonlinear EFT reproduces the full-theory result in the
strong-coupling limit, taking  the process $h h \rightarrow h h$
as an example.
In the full theory, the amplitude for $h h \rightarrow h h  $ is given by
${\cal M}={\cal M}_1 + {\cal M}_2$, where
\begin{align}\label{eq:7.1}
-i\mathcal{M}_1 =&   24 z_1 - 4 d_2^2  
\left[\frac{1}{s_M-M^2} +\frac{1}{t_M-M^2} +\frac{1}{u_M-M^2} \right]   
\end{align}
is the local contribution, from the quartic interaction and 
from $H$-boson exchange, and
\begin{align}\label{eq:7.12}
-i\mathcal{M}_2 =&   -36 d_1^2 \left[    \frac{1}{s_M-m^2} + 
\frac{1}{t_M-m^2} +\frac{1}{u_M-m^2}   \right]     
\end{align}
is the nonlocal term from the exchange of $h$. ${\cal M}_2$ is identical
in the full theory and in the low-energy EFT. We therefore concentrate 
on ${\cal M}_1$ in the following. In the heavy-$H$ limit, we have
\begin{align}\label{eq:7.2}
\frac{1}{s_M-M^2} = - \frac{1}{M^2} \left(1 + \frac{s_M}{M^2}  + \cdots  \right) 
\end{align}
Since the Mandelstam variables satisfy $s_M + t_M + u_M = 4 m^2$, we find
\begin{align}\label{eq:7.3}
-i\mathcal{M}_1 =& 24 z_1 + 12 \frac{d_2^2}{M^2} + 16 \frac{  d_2^2 m^2 }{M^4}  
\end{align}
Fully expanded in the strong-coupling limit, this gives
\begin{align}\label{eq:7.4}
-i\mathcal{M}_1 =& \frac{m^2}{v^2 v^2_s}
\left( 19 s^2 c^2(s v+ cv_s)^2 - 3(s^4 v^2 + c^4 v^2_s)\right)
\end{align}
which coincides with the amplitude from the local $h^4$-term of
the nonlinear EFT in (\ref{fuhvh}).

\section{Linear EFT and comparison with nonlinear EFT}
\label{sec:lin}

We now consider the weakly-coupled limit (\ref{wlimit}) of the singlet model. 
We integrate out the heavy field $S$, while retaining the doublet $\Phi$. 
This is consistent, even though $S$ and $\Phi$ are not the
physical fields. The key point is that the mixing is subleading and
diagonalization is not needed, as opposed to the nonlinear EFT case. 
The resulting Lagrangian can be 
expanded in canonical dimensions. The dominant corrections come from terms 
of dimension six \cite{Grzadkowski:2010es,Buchmuller:1985jz}. For the singlet 
model, this was discussed already in \cite{Gorbahn:2015gxa,Brehmer:2015rna}. 
Starting from \eqref{eq:2.1} and \eqref{eq:2.2}, and rewriting 
$S = (v_{H}+H_s)/\sqrt{2}$, we find
\begin{align}
 \label{eq:5.1}
 \mathcal{L} &= (D^{\mu} \Phi)^{\dag} (D_{\mu} \Phi) +\left(\frac{ \mu_1^2}{2}-
  \frac{\lambda_{3}v_{H}^{2}}{4}\right) \Phi^{\dag} \Phi  - 
  \frac{\lambda_1}{4}  ( \Phi^{\dag} \Phi )^2 \nonumber \\
  &+ \frac{1}{2} \partial^{\mu}  H_s \partial_{\mu} H_s -
 \frac{1}{2} M^2_s H_{s}^{2} \nonumber \\
  & - \frac{\lambda_{3} v_H}{2} \Phi^{\dag} \Phi H_s -  
  \frac{\lambda_{3}}{4} \Phi^{\dag} \Phi H_{s}^{2} - 
 \frac{\lambda_{2}v_H}{4} H_{s}^{3}- \frac{\lambda_{2}}{16} H_{s}^{4}
\end{align}
where we identify
\begin{equation}
  \label{eq:5.2}
\mu_{2}^{2} = M^2_s = \frac{\lambda_2 v_{H}^{2}}{2}
\end{equation}
The equation of motion has the form of \eqref{eq:4.3}, the currents 
$J_{i}$ are now constants and functions of $(\Phi^{\dag} \Phi)$ that can be 
read off from \eqref{eq:5.1}. 
As in the previous section, we solve the equation of motion order 
by order in powers of $1/M_s$. Keeping in mind that 
$v_{H}/M_s = \mathcal{O}(1)$ because of \eqref{eq:5.2}, we find
\begin{equation}\label{solhs}
H_s=-\frac{\lambda_3 v_H}{2 M^2_s} \Phi^\dagger\Phi +
{\cal O}\left(\frac{1}{M^2_s}\right)
\end{equation}
and
\begin{align}
 \label{eq:5.3}
 \mathcal{L} &= (D^{\mu} \Phi)^{\dag} (D_{\mu} \Phi) +\left(\frac{ \mu_1^2}{2}-
 \frac{\lambda_{3}M^2_s}{2\lambda_2}\right)\Phi^{\dag}\Phi -
\left(\frac{\lambda_1}{4}-
 \frac{\lambda_{3}^{2}}{4\lambda_2}\right)( \Phi^{\dag} \Phi )^2 \nonumber \\
  &+ \frac{1}{4}\frac{\lambda_{3}^{2}}{\lambda_2 M^2_s} 
  \partial^{\mu}( \Phi^{\dag} \Phi ) \partial_{\mu}( \Phi^{\dag} \Phi ) + 
  \mathcal{O}\left(\frac{1}{M^4_s}\right)
\end{align}
in agreement with \cite{Gorbahn:2015gxa}. 
Out of the two custodially symmetric scalar operators of dimension six
in the SM, only $\partial^\mu(\Phi^\dagger \Phi)\partial_\mu(\Phi^\dagger \Phi)$
appears. The second operator, $(\Phi^\dagger \Phi)^3$, is absent.\\ 
At low energies, the doublet develops a vev, 
\begin{equation}\label{phihu}
\Phi = \frac{v+h}{\sqrt{2}} U 
\left(\begin{array}{c} 0\\ 1\end{array}\right)
\end{equation}
and $h$ is identified with the light scalar discovered at the LHC. In the 
broken phase, the dimension-6 correction in (\ref{eq:5.3}) translates to
\begin{equation}
  \label{eq:5.4}
 \mathcal{L}_{NLO} = \alpha^2\, \frac{1}{2} 
\partial_{\mu} h \partial^{\mu}h \;  \left(1+\frac{h}{v}\right)^{2}
\end{equation}
with
\begin{equation}\label{alphadef}
\alpha\equiv \frac{\lambda_3}{\lambda_2}\frac{v}{v_H} \dot= \chi
\end{equation}
To first order in $v/v_H$, $\alpha$ is equal to the mixing angle
$\chi$ in (\ref{eq:2.5}). We remove the term in (\ref{eq:5.4}) 
by a field redefinition of $h$ \cite{Buchalla:2013rka}. 
The complete EFT Lagrangian including terms of order $1/M^2_s$ then becomes
(with ${\cal L}_0$ from (\ref{lsm0}))
\begin{align}
\label{eq:5.5}
\mathcal{L} &= \mathcal{L}_0 + 
\frac{1}{2} \partial_{\mu} h \partial^{\mu}h \nonumber \\
&+\frac{v^2}{4}\langle D_\mu U^\dagger D^{\mu} U \rangle \, 
\left( 1+(2-\alpha^{2})\frac{h}{v} +(1-2\alpha^{2})\left(\frac{h}{v}\right)^2 -
\frac{4}{3} \alpha^{2} \left(\frac{h}{v}\right)^3-\frac{\alpha^{2}}{3} 
\left(\frac{h}{v}\right)^4\right)
\nonumber \\
&-\frac{m^2}{2} h^2 - \frac{m^2 v^2}{2}\left[ 
\left( 1-\frac{3}{2}\alpha^{2}\right)\left(\frac{h}{v}\right)^3 +
\left(\frac{1}{4}- \frac{25}{12}\alpha^{2}\right)
\left(\frac{h}{v}\right)^4 -\alpha^{2}\left(\frac{h}{v}\right)^5- 
\frac{\alpha^{2}}{6}\left(\frac{h}{v}\right)^6\right]\nonumber\\
& - v \left[ \bar q Y_u U P_+ \textfrak{r} + \bar q Y_d U P_- \textfrak{r} + 
\bar l Y_e U P_-\eta + 
{\rm h.c.}\right]\times
\nonumber\\
& \hspace*{5cm} \times\left( 1+(1-\frac{\alpha^{2}}{2})\frac{h}{v} -
\frac{\alpha^{2}}{2}\left(\frac{h}{v}\right)^2 -
\frac{\alpha^{2}}{6}\left(\frac{h}{v}\right)^3\right)
\end{align}
We observe that all Higgs couplings in (\ref{eq:5.5}) are reduced
with respect to their Standard-Model values.

\vspace*{0.7cm}

In Section~\ref{sec:nonlin} we performed the matching of the SM extended 
by a heavy scalar singlet to the leading order of the nonlinear EFT by 
integrating out the heavy degree of freedom at tree level. 
We showed that such a low-energy 
EFT is the result of integrating out the heavy field when the theory 
approaches a strongly-coupled regime.    
In the present section, we carried out a matching of the 
theory to the linear EFT through operators of dimension six by integrating out 
the heavy scalar in the weakly-coupled regime. 
We now compare these two scenarios further, based on the
discussion in Section~\ref{sec:model}.   

As stressed previously, the character of the low-energy EFT is dictated by 
the underlying dynamics. In the model at hand, the difference between weak and 
strong coupling, and the respective EFTs, is connected to the size of the 
parameters $\xi$ and $\omega=\sin^2\chi$, 
where $\omega$ quantifies the admixture of the doublet 
and singlet components in the physical scalar fields. When the theory 
approaches the strongly-coupled regime, we have $\xi$, $\omega={\cal O}(1)$. 
The heavy mass eigenstate that is integrated out then has a significant 
doublet component (see also \cite{Burgess:2014lza} for a similar observation 
in a different context). In the weakly-coupled regime the mixing angle $\chi$
shows instead a typical decoupling behavior between 
the two mass scales of the theory, $\omega \sim v^2/f^2$, and
$\xi$, $\omega\ll 1$.   

\begin{figure}[th!]
\centering
\includegraphics[width=8.2cm]{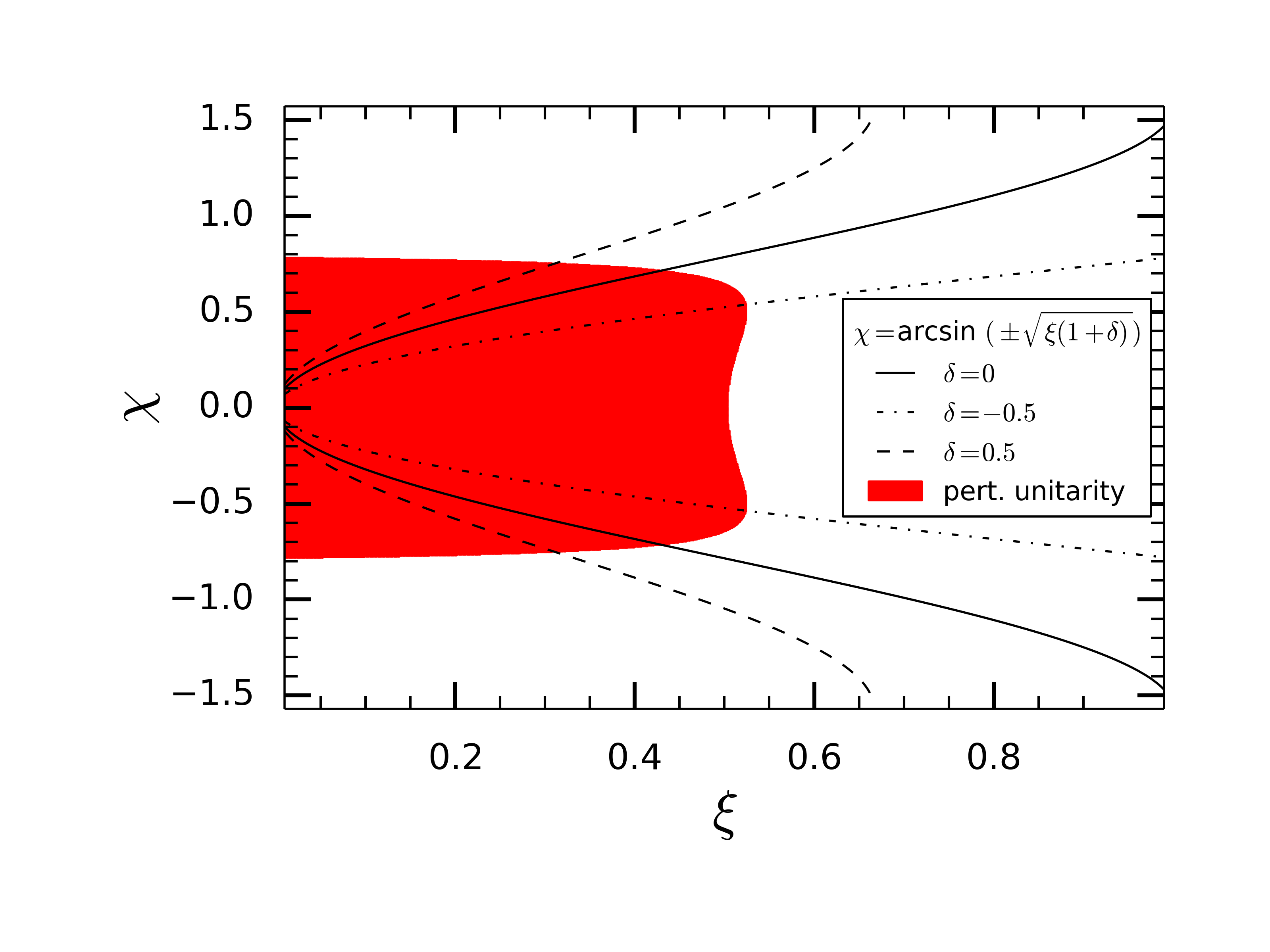}
\hspace{-0.9cm}
\includegraphics[width=8.2cm]{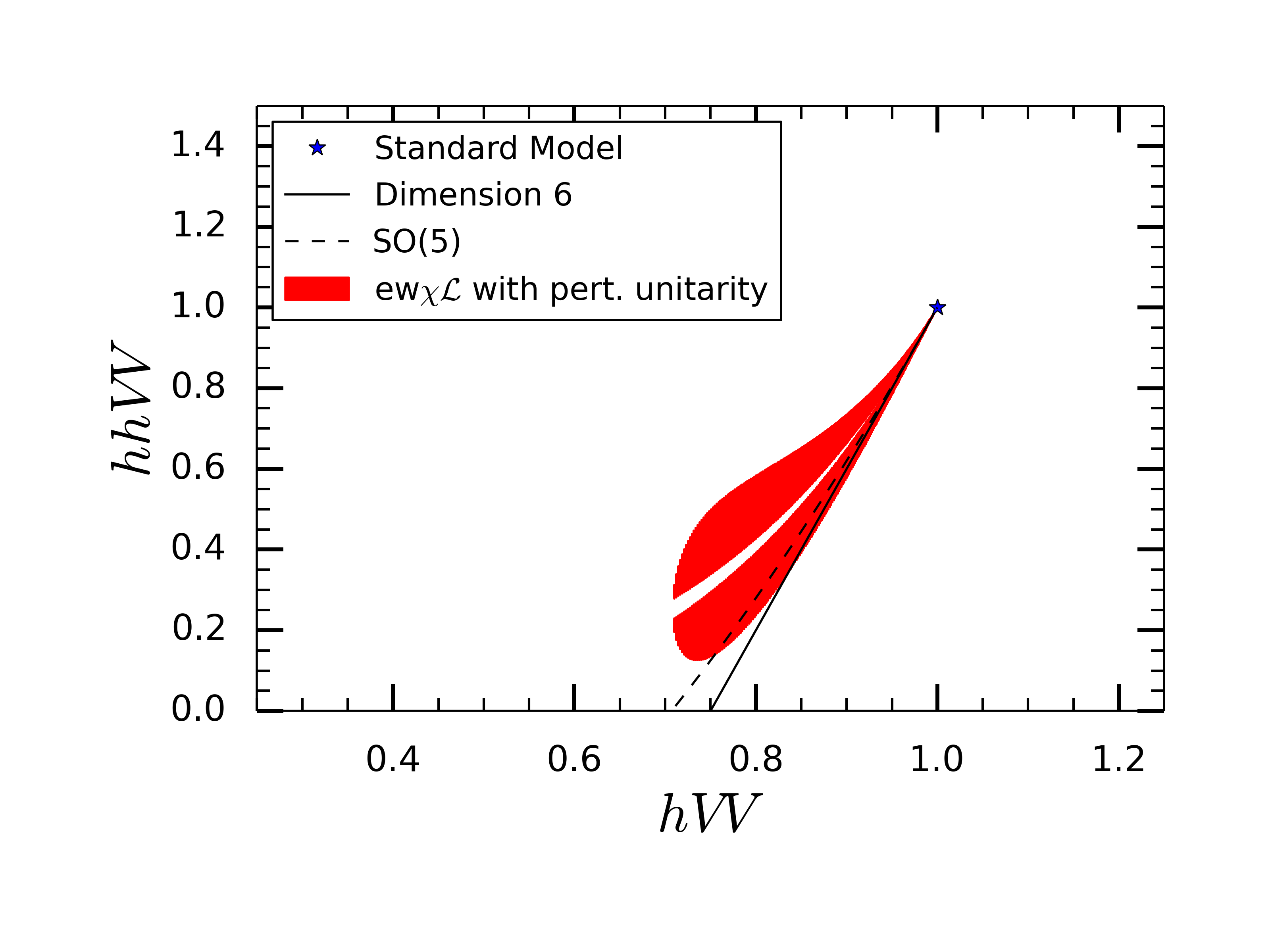}  \vspace{-1.0cm}
\caption{\small   Left:  Allowed values in the plane $\{\xi, \chi\}$ for 
$M=1$~TeV when imposing tree-level perturbative unitarity conditions in the 
full theory. The lines correspond to the $SO(5)$ limit and perturbations
around it.
Right: Illustration of the resulting Higgs couplings to massive 
vector bosons and the decorrelation from the linear EFT at dimension six.}
\label{fig:plots}
\end{figure}

These considerations clarify the connection between the two scenarios 
for the low-energy EFT. Starting from the nonlinear EFT, 
and taking the limit of a small mixing angle, we recover the linear expansion 
of the EFT. In fact, expanding the leading order nonlinear effective 
Lagrangian derived in Section~\ref{sec:nonlin} 
through $\mathcal{O}(\omega,\xi)$, 
we reproduce the results of the linear expansion given in (\ref{eq:5.5}).
We emphasize that in the limit of small mixing angle the linear EFT through
operators of dimension six provides, in particular, a correct description of 
the leading mixing effects in single-Higgs and multiple-Higgs interactions.  

When the mixing angle $\chi\sim v/f$ becomes large, 
indicating the onset of a strongly-coupled regime, the linear expansion 
starts to fail and the nonlinear character of the low-energy EFT becomes 
manifest.  In this scenario, the deviations of the Higgs properties from the 
SM are generically of $\mathcal{O}(1)$ and correspond to a resummation in 
$\chi$ and $\xi$. Another typical feature of the 
nondecoupling behavior is the decorrelation between the linear and quadratic 
Higgs couplings to massive vector bosons~\cite{Contino:2013gna}.   
The latter are linearly correlated in the linear EFT at dimension six,
as seen from (\ref{eq:5.5}). Such a correlation is not present in the 
leading order of the nonlinear EFT, as shown 
in (\ref{fuhvh}). In order to illustrate the size of such effects within the 
perturbative domain of the full theory, we fix $M=1$~TeV and scan the 
remaining $(\xi, \chi)$ parameter space of the model, imposing tree-level 
perturbative unitarity bounds for all two-to-two processes involving 
$\{W^+ W^-, ZZ, hh, h H, HH\}$~\cite{Pruna:2013bma}. Figure~\ref{fig:plots} 
shows the resulting linear and quadratic Higgs coupling to massive 
vector bosons (conveniently normalized) in the nonlinear EFT from this scan. 
For comparison, it also shows the correlations obtained between these two 
couplings near the $SO(5)$-symmetric limit and within the linear expansion 
at dimension six.
The gap between the two regions in Figure~\ref{fig:plots} (right) originates
from the regions of parameter space in which $\chi$ is close to zero.
The Higgs couplings have to be close to their SM values in this case.

\section{\boldmath 
Leading order nonlinear Lagrangian up to 
$\mathcal{O}(\xi^{2})$}
\label{sec:xi2}

In the present section we consider the general, model-independent
electroweak chiral Lagrangian. We assume that a decoupling limit exists,
in which the chiral Lagrangian reduces to the renormalizable Standard Model.
Deviations from this limit are parametrized by a quantity $\xi\equiv v^2/f^2$,
defined in terms of the scale $f$ of the new strong dynamics. 
Corrections at ${\cal O}(\xi^n)$ enter through operators of canonical
dimension $2n+4$ \cite{Buchalla:2013rka,Buchalla:2014eca}. We connect this 
general picture with the singlet model at the end of the section. 

We start from the electroweak chiral Lagrangian at leading order and perform 
the matching to the linear expansion up to $\mathcal{O}(\xi^{2})$ along the 
lines of \cite{Buchalla:2013rka,Buchalla:2014eca}. We neglect terms of 
$\mathcal{O}(\xi/16\pi^{2})$ from higher orders in the chiral expansion, 
which can be justified as long as $\xi^2 \gg \xi/(16 \pi^2)$. 
We write the Higgs sector of 
the LO effective Lagrangian in the dimensional expansion as
\begin{equation}
  \label{eq:xiL}
  \mathcal{L} = \mathcal{L}_{4}+\mathcal{L}_{6}+\mathcal{L}_{8}
\end{equation}
Here $\mathcal{L}_{d}$ contains those operators of chiral dimension 2
that have canonical dimension $d$. The corresponding terms can be 
expressed in terms of the Goldstone matrix $U$ and the Higgs singlet $h$.

At chiral dimension 2 and canonical dimension 4, we have the Higgs
sector of the Standard Model: 
\begin{eqnarray}\label{eq:xi0}
 \mathcal{L}_{4}&=& \frac{1}{2}\partial_{\mu}h\partial^{\mu}h + 
 \frac{\mu^{2}v^{2}}{2}\left(1+\frac{h}{v}\right)^{2} -
\frac{\lambda v^{4}}{8}\left(1+\frac{h}{v}\right)^{4} \nonumber \\
     &-& v\left(1+\frac{h}{v}\right)\bar{\Psi}Y_{\Psi}^{(0)}U P \Psi + 
\frac{v^{2}}{4}\langle D_{\mu}U^\dagger D^{\mu}U\rangle
\left(1+\frac{h}{v}\right)^{2}
\end{eqnarray}
The SM at dimension 6 \cite{Grzadkowski:2010es} contains exactly three
operators that contribute with chiral dimension 2. These are
$\kappa^2 (\Phi^\dagger\Phi)^3$ (including two weak couplings $\kappa$),
$\partial_\mu(\Phi^\dagger\Phi)\partial^\mu(\Phi^\dagger\Phi)$, and the
modified Yukawa terms, here generically written as
$\bar\Psi_L Y\Phi\Psi_R\, \Phi^\dagger\Phi$.
This gives
\begin{equation}\label{ll6}
\mathcal{L}_{6} = -\frac{\lambda a_1 v^{4}}{12}\xi\left(1+\frac{h}{v}\right)^{6}
+\frac{a_2}{2}\xi \partial_{\mu}h\partial^{\mu}h 
 \left(1+\frac{h}{v}\right)^{2}-v\xi\bar{\Psi}\hat{Y}_{\Psi}^{6}U P \Psi
  \left(1+\frac{h}{v}\right)^{3}
\end{equation}
In a similar way, we construct all operators of canonical dimension 8
and chiral dimension 2 and obtain
\begin{equation}\label{ll8}
\mathcal{L}_{8} = -\frac{\lambda b_1v^{4}}{16}\xi^{2}
  \left(1+\frac{h}{v}\right)^{8}+
\frac{b_2}{2}\xi^{2} \partial_{\mu}h\partial^{\mu}h 
\left(1+\frac{h}{v}\right)^{4}-v\xi^{2}\bar{\Psi}\hat{Y}_{\Psi}^{8}U P \Psi
 \left(1+\frac{h}{v}\right)^{5}
\end{equation}
We define $a_{1}$ and $b_{1}$ with an additional factor 
of $\lambda$ to obtain a convenient normalization. The Lagrangian 
of \eqref{eq:xiL} has to be matched to the leading order chiral 
Lagrangian. In order for the kinetic term to be of the form 
$\partial_{\mu}h\partial^{\mu}h/2$, without any other factors, we have to 
redefine $h$:
\begin{eqnarray}\label{eq:xi1}
  h\rightarrow h \Big\lbrace 1-\frac{\xi}{2} a_{2} &&\left(1+\frac{h}{v}+
\frac{h^2}{3 v^2}\right)+\xi ^2 a_{2}^2 \left(\frac{3}{8}+\frac{h}{v}+
\frac{13}{12} \left(\frac{h}{v}\right)^2+\frac{13}{24} 
\left(\frac{h}{v}\right)^3+\frac{13}{120} 
\left(\frac{h}{v}\right)^4\right) \nonumber \\ 
&&  -\xi ^2 b_{2} \left(\frac{1}{2}+\frac{h}{v}+\left(\frac{h}{v}\right)^2 +
\frac{1}{2} \left(\frac{h}{v}\right)^3+\frac{1}{10} 
\left(\frac{h}{v}\right)^4\right)\Big\rbrace + {\cal O}(\xi^3)
\end{eqnarray}

The parameter $v$ describes the physical vev. We find it by 
requiring the linear term in the potential (after the redefinition above) 
to vanish. We find:
\begin{equation}\label{eq:xi2}
v =\sqrt{\frac{2\mu^{2}}{\lambda}}\left(  1-\frac{a_{1}}{2}\xi + 
\frac{\xi^{2}}{2}\left(\frac{3a_{1}^{2}}{4}-b_{1}\right)+
\mathcal{O}\left(\xi^3\right) \right)
\end{equation}
The quadratic term of the potential should be given by the physical 
Higgs mass $m$. This condition, together with \eqref{eq:xi2} enables us 
to express the bare quantities $\mu$ and $\lambda$ of \eqref{eq:xi0} 
in terms of the physical quantities $v$ and $m$, and the 
coefficients $a_i$, $b_i$:
\begin{align}
  \begin{aligned}
    \label{eq:xi3}
 \mu^{2} &= \frac{m^2}{2}
\left(1+\xi (a_{2}-a_{1})+\xi^{2}(2a_{1}^{2}-a_{1}a_{2}-2 b_{1}+b_{2}) +
 \mathcal{O}\left(\xi^3\right) \right)\\
\lambda &= \frac{m^2}{v^2}
  \left(1+\xi \left(a_{2}-2a_{1}\right)+\xi^2 
\left(4 a_{1}^{2}-2 a_{1}a_{2}-3b_{1}+b_{2}\right)+
 \mathcal{O}\left(\xi^3\right)\right)
  \end{aligned}
\end{align}
The Lagrangian then acquires the following form:
\begin{align}
  \begin{aligned}
    \label{eq:xi5}
    \mathcal{L} &= \frac{1}{2}\partial_{\mu}h\partial^{\mu}h -V(h) + 
\frac{v^{2}}{4}\langle D_{\mu}U^\dagger D^{\mu}U\rangle
\left(1+F_{U}(h)\right) -v\bar{\Psi}\left(Y_{\Psi}+\sum_{n=1}^{5}Y_{\Psi}^{(n)}
\left(\tfrac{h}{v}\right)^{n}\right)U P \Psi\\
  \end{aligned}
\end{align}
with
\begin{eqnarray}\label{eqxiexp}
    V(h) &=& \frac{1}{2} m^2 h^2  \nonumber \\ 
    &+&\frac{1}{2} m^2 v^2 \left[\left(1+\xi  \left(\tfrac{4}{3} a_{1}-
\tfrac{3}{2} a_{2}\right)+\xi^2\left(-\tfrac{2}{3} a_{1} a_{2}+
\tfrac{15}{8}a_{2}^2+4 b_{1}-\tfrac{5}{2} b_{2}-
\tfrac{8}{3}a_{1}^{2}\right)\right) \left(\frac{h}{v}\right)^{3}\right. 
\nonumber \\
    &+& \left(\tfrac{1}{4} +\xi  \left(2 a_{1}-\tfrac{25}{12} a_{2}\right)+
\xi^2 \left(-4 a_{1} a_{2}+\tfrac{11}{2} a_{2}^2+8 b_{1}-
\tfrac{21}{4} b_{2} -4a_{1}^{2}\right)\right) \left(\frac{h}{v}\right)^{4}
\nonumber \\
    &+& \left(\xi  (a_{1}-a_{2})+ 
\xi^2 \left(-\tfrac{37}{6}a_{1} a_{2}-2 a_{1}^{2}+
\tfrac{13}{2} a_{2}^2+7 b_{1}-5 b_{2}\right)\right) \left(\frac{h}{v}\right)^{5}
\nonumber \\
    &+&\left(\tfrac{\xi}{6} (a_{1}-a_{2})+\xi^2 
\left(-\tfrac{25}{6} a_{1} a_{2}-\tfrac{1}{3} a_{1}^{2}+\tfrac{176}{45} a_{2}^2+
\tfrac{7}{2} b_{1}-\tfrac{27}{10} b_{2}\right)\right) 
\left(\frac{h}{v}\right)^{6}  \nonumber \\
    &+& \xi^2 \left(-\tfrac{4}{3} a_{1} a_{2}+\tfrac{6}{5} a_{2}^2+b_{1}-
\tfrac{4}{5} b_{2}\right)  \left(\frac{h}{v}\right)^{7} \nonumber \\
&+&\left. \tfrac{\xi^2}{8} 
\left(-\tfrac{4}{3} a_{1} a_{2}+\tfrac{6}{5} a_{2}^2+b_{1}-
\tfrac{4}{5} b_{2}\right) \left(\frac{h}{v}\right)^{8}\right]
\end{eqnarray}
\begin{eqnarray}\label{fuhxi}
    F_{U}(h) &=&\left(2-a_{2} \xi +
 \xi^2 \left(\tfrac{3}{4} a_{2}^2-b_{2}\right)\right)\left(\frac{h}{v}\right)+ 
\left(1-2 a_{2} \xi+\xi ^2 \left(3 a_{2}^2-3 b_{2}\right)\right)
 \left(\frac{h}{v}\right)^{2} \nonumber \\
    &+&\left(-\xi\tfrac{4}{3} a_{2}+\xi^2 
\left(\tfrac{14}{3} a_{2}^2-4 b_{2}\right)\right)\left(\frac{h}{v}\right)^{3}+ 
\left(-\xi \tfrac{a_{2}}{3}+\xi^2 
\left(\tfrac{11}{3} a_{2}^2-3 b_{2}\right)\right)\left(\frac{h}{v}\right)^{4}
\nonumber \\
    &+&\xi^2 \left(\tfrac{22}{15} a_{2}^2-\tfrac{6}{5} b_{2}\right)
\left(\frac{h}{v}\right)^{5}+\tfrac{\xi^2}{6} \left(\tfrac{22}{15} a_{2}^2-
\tfrac{6}{5} b_{2}\right)\left(\frac{h}{v}\right)^{6} 
\end{eqnarray}
\begin{eqnarray} \label{yhxi}
\sum_{n=1}^{5}Y_{\Psi}^{(n)}\left(\tfrac{h}{v}\right)^{n} &=&\left(Y_{\Psi}+\xi  
\left(2 \hat{Y}_{\Psi}^{6}-\tfrac{a_{2}}{2} Y_{\Psi}\right)+
\xi^2 \left(\tfrac{3}{8} a_{2}^2 Y_{\Psi}-a_{2} \hat{Y}_{\Psi}^{6}-
\tfrac{b_{2}}{2} Y_{\Psi}+4 \hat{Y}_{\Psi}^{8}\right)\right)\frac{h}{v} 
\nonumber \\
    &+&\left(\xi  \left(3 \hat{Y}_{\Psi}^{6}-\tfrac{a_{2}}{2} Y_{\Psi}\right)+
\xi^2 \left(a_{2}^2 Y_{\Psi}-4 a_{2} \hat{Y}_{\Psi}^{6}-b_{2} Y_{\Psi}+
10 \hat{Y}_{\Psi}^{8}\right)\right)\left(\frac{h}{v}\right)^{2}
\nonumber \\
    &+&\left(\tfrac{\xi}{3}  
\left(3 \hat{Y}_{\Psi}^{6}-\tfrac{a_{2}}{2} Y_{\Psi}\right)+\xi^2 
\left(\tfrac{13}{12} a_{2}^2 Y_{\Psi}-
\tfrac{29}{6} a_{2} \hat{Y}_{\Psi}^{6}-b_{2} Y_{\Psi}+
10 \hat{Y}_{\Psi}^{8}\right)\right)\left(\frac{h}{v}\right)^{3}
\nonumber \\
    &+& \xi^2 \left(\tfrac{13}{24} a_{2}^2 Y_{\Psi}-
\tfrac{5}{2} a_{2} \hat{Y}_{\Psi}^{6}-\tfrac{b_{2}}{2} Y_{\Psi}+
 5 \hat{Y}_{\Psi}^{8}\right)\left(\frac{h}{v}\right)^{4} \nonumber \\
    &+& \tfrac{\xi^2}{5} \left(\tfrac{13}{24} a_{2}^2 Y_{\Psi}-
\tfrac{5}{2} a_{2}\hat{Y}_{\Psi}^{6}-\tfrac{b_{2}}{2} Y_{\Psi}+
5\hat{Y}_{\Psi}^{8}\right)\left(\frac{h}{v}\right)^{5}
\end{eqnarray}
where
\begin{equation}\label{yy068}
Y_\Psi = Y_{\Psi}^{(0)} + \xi \hat{Y}_{\Psi}^{6} + \xi^2  \hat{Y}_{\Psi}^{8}
\end{equation}

Comparing (\ref{eqxiexp}), (\ref{fuhxi}) and (\ref{yhxi}) with the results 
for $F_U(h)$, $V(h)$ and the Yukawa terms in the singlet model, 
displayed in (\ref{fuhvhso5}) and (\ref{yukhso5}), we find agreement
to second order in $\xi$ 
with $a_1=b_1=\hat Y^6_\Psi=\hat Y^8_\Psi=0$, $a_2=b_2=1$, $Y_\Psi=Y_f$.

In relation to our previous discussion we make the following observations. 
Since (\ref{eq:5.5}) contains contributions through dimension six, it induces 
a pattern of coefficients that is expected from the $\mathcal{O}(\xi)$ 
expansion of the chiral Lagrangian. Indeed, (\ref{eq:5.5})  can be obtained 
from (\ref{eqxiexp}), (\ref{fuhxi}) and (\ref{yhxi}) by neglecting terms of 
$\mathcal{O}(\xi^2)$ and identifying $a_2 \xi= \alpha^2$.   
This result could have been anticipated also from the analysis in 
\cite{Buchalla:2014eca}. The decorrelation between the linear and quadratic 
Higgs couplings to massive vector bosons appears at dimension eight. 
This is in agreement with the discussion in \cite{Contino:2013gna}.  
Additional correlations at the $\mathcal{O}(\xi)$ level 
are also present in the Yukawa sector ($h^{2}$ and $h^3$ couplings) and in 
the scalar potential ($h^{5}$ and $h^6$ couplings), though these seem less 
interesting phenomenologically.

\section{Conclusions}
\label{sec:concl}

We have studied a simple extension of the Standard Model 
where new physics is limited to a heavy real scalar singlet endowed with 
a $Z_2$ symmetry. This model has been used in the past, e.g., for searches 
of dark matter. Here we use it as a (UV-complete) toy model to 
illustrate, by explicit construction, how the different effective field 
theories at the electroweak scale, the so-called SM-EFT and EWChL, arise. 
These two EFTs possess the same degrees of freedom and symmetries, yet they 
have very different systematics: SM-EFT is an expansion in canonical dimensions 
while EWChL is an expansion in loops or chiral dimensions. 
The toy model allows us to show in a transparent way why this difference 
in power counting occurs, and helps to substantiate by way of example 
a number of statements about both EFTs.
\begin{itemize}
\item {\emph{Dynamics of the EFTs}}. The model depends on three free 
parameters: the heavy mass $M$, the mixing angle $\omega$ and the vev ratio 
$\xi$. In scenarios where $M$ is large and $\omega,\xi\ll 1$, the heavy scalar 
scales as $H\sim {\cal{O}}(M^{-1})(v+h)^2$ and the resulting 
EFT is organized in inverse powers of $M$ (SM-EFT). In contrast, if 
$\omega,\xi\sim {\cal{O}}(1)$, then $H\sim {\cal{O}}(1) f(h)$, with $f(h)$ 
an (untruncated) function of $h$. This corresponds to a nondecoupling regime 
and the EFT is then organized in chiral dimensions (EWChL). Generically, 
theories that exhibit nondecoupling effects lead to EFTs governed by chiral 
dimensions, while theories with only decoupling effects admit EFTs based on 
an expansion in canonical dimensions.
\item {\emph{Relation between the EFTs}}. The model shows that the choice of 
EFT depends only on the size of the parameters. The transition between EWChL 
and SM-EFT is therefore a smooth one, as can be shown by further expanding 
EWChL for small $\omega,\xi$. This conclusion holds as long as there is a 
well-defined decoupling limit.
\item {\emph{$\xi$ expansion}}. In a bottom-up EFT the $\xi$ dependence is 
hidden in the Wilson coefficients and cannot be determined from power 
counting. One can nevertheless uncover this $\xi$ dependence in EWChL 
starting from operators in SM-EFT~\cite{Buchalla:2014eca}. Here we have 
extended this procedure to the leading-order EWChL at ${\cal{O}}(\xi^2)$ and 
compared it explicitly with the toy model expanded at the same order. 
We find a consistent matching, which validates the procedure adopted 
in~\cite{Buchalla:2014eca}.
\item {\emph{Naturalness}}. The toy model at hand 
(in the nondecoupling regime) admits an embedding into an $SO(5)$ model 
spontaneously broken down to $SO(4)$. In that case, and if explicit breaking 
of the $SO(5)$ symmetry is small, the Higgs can be interpreted as a 
pseudo-Goldstone boson and is therefore naturally light, $m\sim f$. 
Its precise value cannot be computed in perturbation theory unless one 
assumes that $M/(4\pi f)\lesssim 1$ makes the loop expansion sufficiently 
convergent. Away from the $SO(5)$ limit, fine-tuning is required to build a 
hierarchy between the Higgs and the heavy scalar.
\end{itemize}

\section*{Acknowledgements}

This work was performed in the context of the ERC Advanced Grant 
project FLAVOUR (267104) and was supported in part by the DFG grant 
BU 1391/2-1, the DFG cluster of excellence EXC 153 'Origin and Structure of 
the Universe' and the Munich Institute for Astro- and Particle Physics (MIAPP). 
A.C. is supported by a Research Fellowship
of the Alexander von Humboldt Foundation.

\appendix

\numberwithin{equation}{section} 

\section{Exact solution for $H_0(h)$}
\label{sec:exh0}

We integrate out the heavy field $H$ with mass $M$ at tree level by 
solving its equation of motion. In the strong-coupling limit (\ref{slimit})
the leading-order term $H_0(h)$, of order ${\cal O}(1)$ in the $1/M^2$ 
expansion, follows from solving the equation of motion at ${\cal O}(M^2)$. 
We achieved this in Section \ref{sec:nonlin} through a series expansion 
of $H_0$ in powers of $h$. Here we obtain an exact, analytic solution
for the function $H_0(h)$.

Retaining only the terms of order $M^2$, sufficient for the computation of
$H_0(h)$, the Lagrangian of the full singlet model simplifies to
(see (\ref{eq:2.1}))
\begin{equation}\label{lm2} 
{\cal L}_{M}=\frac{\mu^2_1}{2}\phi^2 + \frac{\mu^2_2}{2} S^2 
-\frac{\lambda_1}{4}\phi^4 -\frac{\lambda_2}{4} S^4 
-\frac{\lambda_3}{2} \phi^2 S^2
\end{equation}
where $\phi^2\equiv \Phi^\dagger\Phi$ and
\begin{equation}\label{lamm2}
\lambda_1=\frac{2M^2}{f^2}\frac{\omega}{\xi}\, ,\qquad
\lambda_2=\frac{2M^2}{f^2}\frac{1-\omega}{1-\xi}\, ,\qquad
\lambda_3=\frac{2M^2}{f^2}\sqrt{\frac{\omega(1-\omega)}{\xi(1-\xi)}}
\end{equation}
\begin{equation}\label{mum2}
\mu^2_1=
M^2\left(\omega+\sqrt{\frac{\omega}{\xi}}\sqrt{(1-\omega)(1-\xi)}\right)\, ,
\qquad
\mu^2_2=M^2\left(1-\omega+\sqrt{\frac{1-\omega}{1-\xi}}\sqrt{\omega\xi}\right)
\end{equation}
Expanding $\phi$ and $S$ around their vevs and using (\ref{eq:2.4}),
we write
\begin{eqnarray}\label{phishh}
\phi &=& \frac{1}{\sqrt{2}}(f\sqrt{\xi} +\sqrt{1-\omega}h+\sqrt{\omega}H)
\nonumber\\
S &=& \frac{1}{\sqrt{2}}(f\sqrt{1-\xi} -\sqrt{\omega}h+\sqrt{1-\omega}H)
\end{eqnarray}
Defining next
\begin{equation}\label{r2def}
R^2\equiv \sqrt{\frac{\omega}{\xi}} \phi^2 +\sqrt{\frac{1-\omega}{1-\xi}}S^2
\end{equation}
the Lagrangian (\ref{lm2}) becomes
\begin{equation}\label{lm2r2}
{\cal L}_{M} =
\frac{M^2}{2}\left(\sqrt{\omega\xi}+\sqrt{(1-\omega)(1-\xi)}\right) R^2
-\frac{M^2}{2f^2} R^4
\end{equation}
The resulting equation of motion for $H$ reads
\begin{equation}\label{hm2eom}
\frac{\partial {\cal L}}{\partial H}=
\frac{\partial {\cal L}}{\partial R^2}\ \frac{\partial R^2}{\partial H}=0
\end{equation}
The relevant solution $H_0(h)$, inserted back into ${\cal L}$, describes the  
effect of integrating out $H$ at tree level. This is equivalent to matching
all possible tree diagrams with internal $H$ lines 
to an effective Lagrangian for $h$. 

The Lagrangian for $H$ has the form of (\ref{eq:4.1}).
A diagram with only internal $H$ lines contains, in general,
a number $V_n$ of vertices $J_n H^n$ ($n=1,\ldots, 4$), $P$ $H$-field 
propagators, and $L$ loops.
Combining the well-known topological identities
\begin{align}\label{pvil}
2P &= V_1 + 2 V_2 + 3 V_3 + 4 V_4 \nonumber\\
L &= P-(V_1+V_2+V_3+V_4) + 1 
\end{align}
for the number of $H$-lines attached to vertices and the number of loops,
respectively, one obtains
\begin{equation}\label{llvi}
L = V_4 +\frac{V_3-V_1}{2}+1
\end{equation}
For tree diagrams ($L=0$), this implies
\begin{equation}\label{v134}
V_1 = V_3 + 2 V_4 + 2
\end{equation}
Since $V_3$, $V_4\geq 0$, we find $V_1\geq 2$. This means that the effective
Lagrangian, obtained from (\ref{eq:4.1}) by integrating out $H$ at tree level,
has to start at order $(J_1)^2$. Equivalently, the solution of the equation 
of motion for $H$ has to start at ${\cal O}(J_1)$. To order $M^2$, relevant
for $H_0$, this implies that $H_0(h)={\cal O}(h^2)$.


This consideration eliminates the solution for $H(h)$ of
\begin{equation}\label{drdh}
0=\frac{\partial R^2}{\partial H}=f+\sqrt{\omega(1-\omega)}
\left(\sqrt{\frac{\omega}{\xi}}-\sqrt{\frac{1-\omega}{1-\xi}}\right) h +
\left(\omega\sqrt{\frac{\omega}{\xi}}+
(1-\omega)\sqrt{\frac{1-\omega}{1-\xi}}\right) H
\end{equation}
and one of the solutions of the equation $\partial{\cal L}/\partial R^2=0$,
quadratic in $H$, which can also be written as
\begin{equation}\label{dldh} 
R^2 =\frac{f^2}{2}\left(\sqrt{\omega\xi}+\sqrt{(1-\omega)(1-\xi)}\right)
\end{equation}
The remaining solution of (\ref{dldh}) is
\begin{align}\label{h0exact}
H_0(h) &=\frac{f+\left(\sqrt{\frac{\omega^2(1-\omega)}{\xi}}-
\sqrt{\frac{\omega(1-\omega)^2}{1-\xi}}\right)h}{
\sqrt{\frac{\omega^3}{\xi}}+\sqrt{\frac{(1-\omega)^3}{1-\xi}}}
\nonumber\\
& \hspace*{1.5cm} \times\left[\sqrt{1-\frac{
\left(\sqrt{\frac{\omega^3}{\xi}}+\sqrt{\frac{(1-\omega)^3}{1-\xi}}\right)
\left(\sqrt{\frac{\omega(1-\omega)^2}{\xi}}+
\sqrt{\frac{\omega^2(1-\omega)}{1-\xi}}\right)h^2}{
\left(f+\left(\sqrt{\frac{\omega^2(1-\omega)}{\xi}}-
\sqrt{\frac{\omega(1-\omega)^2}{1-\xi}}\right)h\right)^2}} -1\right]
\end{align}
As expected, $H_0(h)={\cal O}(h^2)$.
All coefficients of $h^n$ in $H_0(h)$ are polynomial in
$\sqrt{\omega}=\sin\chi$ and $\sqrt{1-\omega}=\cos\chi$.

Expanded in powers of $h$, (\ref{h0exact}) agrees with the result for $H_0$
obtained in Section \ref{sec:nonlin} through order $h^5$.
In the $SO(5)$ limit, where $\omega=\xi$, (\ref{h0exact}) becomes
\begin{equation}\label{h0so5}
H_0=f\left[\sqrt{1-\frac{h^2}{f^2}}-1\right]
\end{equation}

As a byproduct of our derivation, we can show explicitly that the
terms of order $M^2$ cancel out in the effective Lagrangian, as it
has to be the case.
This property is not immediately obvious from the full theory
in (\ref{eq:3.3}), where the coefficients carry ${\cal O}(M^2)$
contributions.
From (\ref{dldh}) we see that the solution for $H_0$ 
fulfills $R^2(h, H_0(h))=const$. Therefore, when $H_0(h)$ is inserted
back into  (\ref{lm2r2}), the $M^2$-terms in the Lagrangian reduce
to a field-independent constant. This demonstrates the absence of
a nontrivial ${\cal O}(M^2)$ piece in the effective theory.

\section{The scalar effective potential to one loop}
\label{sec:veff}

We consider the one-loop effective potential
of the scalar sector defined in (\ref{eq:2.1}) and (\ref{eq:2.2}),
when the heavy field is integrated out.
The result illustrates the parametric impact of radiative
corrections within the model, in particular in the strong-coupling limit.

We start from the scalar Lagrangian of the model in terms of
the mass eigenstates, given by 
\begin{equation}\label{lscal}
{\cal L}=-\frac{1}{2}h\partial^2 h -\frac{1}{2}H\partial^2 H
-\frac{1}{2} m^2 h^2 -\frac{1}{2}M^2 H^2 - V_{34}
\end{equation}
where $V_{34}$ are the cubic and quartic terms of $V(h,H)$ in (\ref{eq:3.3}).
Following the background-field methods described in \cite{Dittmaier:1995cr},
we split the fields into a background component, denoted by a hat, and
a fluctuating part
\begin{equation}\label{hhshift}
h\to \hat h+ h\, ,\qquad H\to \hat H+H
\end{equation}
For the one-loop computation, we need the part of $V_{34}$ quadratic in the 
fluctuating fields. It reads
\begin{equation}\label{v342}
-V_{34,2}=A h^2 + B H^2 + 2 C h H
\end{equation}
where
\begin{align}\label{abcdef}
A =& 3 d_1 \hat h + d_2\hat H+ 6 z_1 \hat h^2+ 3z_2 \hat h \hat H +
   z_3 \hat H^2\nonumber \\
B =& d_3 \hat h + 3 d_4\hat H+  z_3 \hat h^2+ 3z_4 \hat h \hat H +
   6 z_5 \hat H^2\nonumber \\
C =& d_2 \hat h +  d_3\hat H+  \frac{3}{2}z_2 \hat h^2+ 2z_3 \hat h \hat H +
   \frac{3}{2} z_4 \hat H^2
\end{align}
The Lagrangian terms of second order in the fluctuating fields 
$h$ and $H$ then become
\begin{equation}\label{lhh2}
{\cal L}_2 =-\frac{1}{2}h\partial^2 h -\frac{1}{2}H\partial^2 H
-\frac{1}{2} m^2 h^2 -\frac{1}{2}M^2 H^2 + A h^2 + B H^2 + 2 C h H
\end{equation}
We next isolate the dependence on $M^2$ that is still hidden in the
coefficients $d_i$ and $z_i$ in (\ref{abcdef}).
Following Appendix \ref{sec:exh0}, the full $M^2$ dependence takes the
form of ${\cal L}_M$ in (\ref{lm2r2}). The terms of second order in
$h$ and $H$ are
\begin{equation}\label{l2m2}
{\cal L}_{2,M} =
\frac{1}{2}\left.\frac{\partial^2{\cal L}_M}{\partial h^2}\right|\subhat\ h^2 +
\frac{1}{2}\left.\frac{\partial^2{\cal L}_M}{\partial H^2}\right|\subhat\ H^2 +
\left.\frac{\partial^2{\cal L}_M}{\partial h \partial H}\right|\subhat\ h H
\end{equation}
where the subscript \raisebox{-7pt}{$\hat{}$} after an expression indicates 
that its field variables are taken at their background values. 
The second derivatives are
\begin{eqnarray}\label{lm2d}
\frac{\partial^2{\cal L}_M}{\partial h^2} &=&
\frac{\partial^2{\cal L}_M}{(\partial R^2)^2}
\left(\frac{\partial R^2}{\partial h}\right)^2 +
\frac{\partial {\cal L}_M}{\partial R^2}\frac{\partial^2 R^2}{\partial h^2}
\nonumber\\
\frac{\partial^2{\cal L}_M}{\partial H^2} &=&
\frac{\partial^2{\cal L}_M}{(\partial R^2)^2}
\left(\frac{\partial R^2}{\partial H}\right)^2 +
\frac{\partial {\cal L}_M}{\partial R^2}\frac{\partial^2 R^2}{\partial H^2}
\nonumber\\
\frac{\partial^2{\cal L}_M}{\partial h \partial H} &=&
\frac{\partial^2{\cal L}_M}{(\partial R^2)^2}
\frac{\partial R^2}{\partial h} \frac{\partial R^2}{\partial H} +
\frac{\partial {\cal L}_M}{\partial R^2}
\frac{\partial^2 R^2}{\partial h \partial H}
\end{eqnarray}
where $R^2$ is defined in (\ref{phishh}) and (\ref{r2def}).
We then have
\begin{eqnarray}\label{drdhh}
\frac{\partial R^2}{\partial H} &=& f+
\left(\sqrt{\frac{1-\omega}{\xi}}\omega -
  \sqrt{\frac{\omega}{1-\xi}}(1-\omega)\right) h +
\left(\sqrt{\frac{\omega}{\xi}} \omega +
  \sqrt{\frac{1-\omega}{1-\xi}}(1-\omega)\right) H \nonumber\\
\frac{\partial R^2}{\partial h} &=&
\left(\sqrt{\frac{\omega}{\xi}}(1-\omega)+
  \sqrt{\frac{1-\omega}{1-\xi}}\omega\right) h +
\left(\sqrt{\frac{1-\omega}{\xi}} \omega -
  \sqrt{\frac{\omega}{1-\xi}}(1-\omega)\right) H 
\end{eqnarray}
Evaluated with background fields to leading order in $M^2$, 
the expressions in (\ref{lm2d}) simplify because
\begin{equation}\label{dlmhat}
\left.\frac{\partial {\cal L}_M}{\partial R^2}\right|\subhat = 0\, , \qquad
\left.\frac{\partial^2 {\cal L}_M}{(\partial R^2)^2}\right|\subhat = 
-\frac{M^2}{f^2} = const.
\end{equation}
from (\ref{lm2r2}) and (\ref{dldh}).
Defining
\begin{equation}\label{alphabeta}
\alpha\equiv \left(\frac{\partial R^2}{\partial h}\right)\subhat\quad , \qquad
\beta\equiv \left(\frac{\partial R^2}{\partial H}\right)\subhat
\end{equation}
we obtain
\begin{equation}\label{l2mab}
{\cal L}_{2,M} =
-\frac{M^2}{2f^2}\left( \alpha^2 h^2 + \beta^2 H^2 + 2\alpha\beta h H\right)
\end{equation}
This result gives an explicit expression for the $M^2$-dependent 
terms contained in (\ref{lhh2}).

We now write the second-order Lagrangian in (\ref{lhh2}) as
\begin{equation}\label{l2kk}
{\cal L}_2=-\frac{1}{2} (h,H)
\left(
\begin{array}{cc}
\Delta + a & c \\
c & \Delta +b
\end{array}
\right)
\left(
\begin{array}{c}
h\\
H
\end{array}
\right)
\equiv -\frac{1}{2} (h,H) K
\left(\begin{array}{c} h\\ H \end{array} \right)
\end{equation} 
Here
\begin{equation}\label{deldef}
\Delta\equiv \partial^2 + m^2
\end{equation}
and
\begin{equation}\label{abcbar}
a\equiv - 2\bar A +\frac{M^2}{f^2}\alpha^2\, ,\qquad
b\equiv - 2\bar B -m^2  +\frac{M^2}{f^2}\beta^2\, ,\qquad
c\equiv - 2\bar C +\frac{M^2}{f^2}\alpha\beta
\end{equation}
where $\bar A$, $\bar B$ and $\bar C$ are, respectively, the functions $A$,
$B$ and $C$ of (\ref{abcdef}) without the $M^2$-pieces. 
The latter are made explicit in (\ref{abcbar}).

We obtain the one-loop effective action $S_{eff}$ from the path integral 
\begin{equation}\label{seffpath}
\int {\cal D}h\, {\cal D}H\, \exp\left[i\int d^4x\, {\cal L}_2\right]
={\rm Det}\left(i K\, \delta^{(4)}(x-y)\right)^{-1/2} =\exp(i S_{eff})
\end{equation}
It follows that
\begin{equation}\label{sefftrln}
S_{eff}=\frac{i}{2} \ln\left( {\rm Det}(K\, \delta^{(4)}(x-y))\right) =
\frac{i}{2} {\rm Tr}\left( \ln K\, \delta^{(4)}(x-y)\right)
\end{equation}
We write \cite{Dittmaier:1995cr}
\begin{equation}\label{lnkxd}
\ln K \, \delta^{(4)}(x-y) =
\int\frac{d^4p}{(2\pi)^4}\ln K(x,\partial_x)\, e^{ip(x-y)} =
\int\frac{d^4p}{(2\pi)^4}\,  e^{ip(x-y)}\, \ln K(x,\partial_x+ ip)
\end{equation}
and find
\begin{equation}\label{trlnk}
{\rm Tr}\left(\ln K\, \delta^{(4)}(x-y)\right)=
\int d^4x\,\int\frac{d^4p}{(2\pi)^4}\, {\rm tr}\left(\ln K(x,\partial_x + ip)
\right)
\end{equation}
Here the trace ${\rm Tr}$ is taken over both space-time indices and the
matrix $K$, the trace ${\rm tr}$ only over $K$. We use a similar convention
for the determinant symbols ${\rm Det}$ and ${\rm det}$.

Inserting (\ref{trlnk}) into (\ref{sefftrln}), we obtain the 
one-loop effective Lagrangian
\begin{equation}\label{lefflndet}
{\cal L}_{eff}=\frac{i}{2}\int\frac{d^4p}{(2\pi)^4}\, 
{\rm tr}\left(\ln K(x,\partial_x + ip)\right)
=\frac{i}{2}\int\frac{d^4p}{(2\pi)^4}\, 
\ln\left({\rm det}\, K(x,\partial_x + ip)\right)
\end{equation}
where
\begin{equation}\label{detkk}
{\rm det}\, K=
\Delta(\Delta + a + b)\left[1+\frac{ab-c^2}{\Delta(\Delta + a + b)}\right]
\end{equation}

In the following, we specialize to the effective potential with constant 
background fields. The derivatives $\partial_x$ of $K$ in (\ref{lefflndet})
can then be dropped and $\Delta\to -p^2 + m^2$. 
Up to an irrelevant constant, the effective Lagrangian becomes
\begin{equation}\label{leff12}
{\cal L}_{eff} = {\cal L}_{eff,1} + {\cal L}_{eff,2}
\end{equation}
with
\begin{equation}\label{leff1}
{\cal L}_{eff,1} =\frac{i}{2}\int\frac{d^4p}{(2\pi)^4}\, \ln(p^2-(a+b+m^2))
\end{equation}
\begin{equation}\label{leff2}
{\cal L}_{eff,2} =\frac{i}{2}\int\frac{d^4p}{(2\pi)^4}\, 
\sum^\infty_{n=1} \frac{(-1)^{n+1}}{n}
\left(\frac{ab-c^2}{(p^2-m^2)(p^2-(a+b+m^2))}\right)^n
\end{equation}

We assume the model has an $SO(5)$ symmetry in the scalar sector,
which is weakly broken, as discussed at the end of Section~\ref{sec:model}.
With the parameter $\delta=\omega/\xi -1 \ll 1$, we find
\begin{equation}\label{al2be2}
\alpha^2 + \beta^2 = (f+\hat H)^2 + \hat h^2 + {\cal O}(f^2 \delta)
= 2 R^2 + {\cal O}(f^2 \delta)
\end{equation}
The equation of motion (\ref{dldh}) gives $R^2 = f^2/2 + {\cal O}(f^2 \delta)$.
This implies
\begin{equation}\label{albef2}
\alpha^2 + \beta^2 = f^2 + {\cal O}(f^2 \delta)
\end{equation}
The field $\hat H=\hat H(\hat h)$ is understood to be expressed
as a function of $\hat h$ from solving the e.o.m., as shown in 
Appendix~\ref{sec:exh0}. 

We then find for the parameters in (\ref{leff1}) and (\ref{leff2})
\begin{equation}\label{abm2}
a+b+m^2 = M^2 - 2 \bar A - 2\bar B +{\cal O}(M^2 \delta)
=M^2 + {\cal O}(v^2)
\end{equation}
and
\begin{equation}\label{abc2}
a b - c^2 =-\frac{M^2}{f^2}\left( (2\bar B+m^2) \alpha^2 + 2 \bar A \beta^2
-4\bar C \alpha\beta\right) +2\bar A(2\bar B + m^2) - 4 \bar C^2
\end{equation}
The leading term of $a+b+m^2$ in the limit (\ref{slimit}) is just $M^2$,
while the remaining $\hat h$-dependent terms are only of order $v^2$:
$\bar A$ and $\bar B$ are of this order by definition, and 
$M^2\delta ={\cal O}(v^2)$ because of (\ref{v1coup}).
The term $ab-c^2$ has a leading, field-dependent part $\sim M^2$ and
subleading contributions of order $v^2$.
Note that terms of order $M^4$ present in $a b$ and $c^2$ cancel in
the difference.

Using (\ref{abm2}), we rewrite ${\cal L}_{eff,1}$ in (\ref{leff1}), up to
a constant, as  
\begin{equation}\label{leff1sum}
{\cal L}_{eff,1} =\frac{i}{2}\int\frac{d^4p}{(2\pi)^4}\, 
\sum^\infty_{n=1} \frac{(-1)^{n+1}}{n}
\left(\frac{2(\bar A+\bar B) +{\cal O}(M^2\delta)}{p^2-M^2}\right)^n 
\end{equation}

The dominant corrections in (\ref{leff2}) and (\ref{leff1sum}) in the 
strong-coupling limit (\ref{slimit}) arise from the first term in the sums 
with $n=1$. Relative to the tree-level potential in (\ref{fuhvh}), they are
of order $M^2/(16\pi^2 f^2)$, up to logarithms $\ln M/m$.
Further terms are subleading, of order $1/(16\pi^2)\cdot (v^2/M^2)^k$,
with $k\geq 0$.
The one-loop corrections to the effective potential in (\ref{leff2})
and (\ref{leff1sum}) are still divergent, requiring renormalization of
the leading-order parameters.

In the regime of large, but still perturbative couplings, as discussed in
Section~\ref{sec:model}, the parameter $M^2/(16\pi^2 f^2)$ is smaller than
unity and the tree level potential remains a meaningful approximation.
In the case of genuine strong coupling, $M^2/(16\pi^2 f^2)\approx 1$ and the 
loop corrections become as large as the tree-level ones. In this scenario,
the heavy scalar would become a broad resonance and the singlet-extension
of the SM would no longer be calculable and consistent. The coefficients
of the effective Lagrangian would then be arbitrary parameters of order
unity, determined by the underlying, uncalculable strong dynamics.
In the weakly-coupled limit (\ref{wlimit}),  
$M^2/(16\pi^2 f^2)\approx 1/(16\pi^2)$, and the loop corrections
are of the usual perturbative size.

The resulting consistent picture of the one-loop corrections to the 
effective potential in the limit (\ref{slimit}) relies
on the approximate $SO(5)$ symmetry of the scalar model, 
as we see from (\ref{abm2}) and (\ref{abc2}). 
The corresponding role of the light Higgs as a pseudo-Goldstone boson
is also illustrated by considering the limit of an exact $SO(5)$
symmetry. In that case $\delta=r=0$, and we find $R^2=((f+H)^2 + h^2)/2$,
$\alpha=\hat h$, and $\beta=f+\hat H$. The equation of motion then fixes 
$\alpha^2+\beta^2=f^2 = const.$, see (\ref{albef2}).
Since $\bar A=\bar B=\bar C=0$, we also have $ab-c^2=0$ and
$a+b+m^2 = M^2$. This implies ${\cal L}_{eff}=-V_{eff}=const.$, so that no
nontrivial potential for $\hat h$ is generated, in accordance with the
Goldstone theorem.

Finally, we give expressions for the leading one-loop corrections,
of order $M^2/(16\pi^2 f^2)$,
to the effective potential in the strong-coupling limit (\ref{slimit}). 
They come from the $n=1$ terms in (\ref{leff2}) and (\ref{leff1sum})
and read in dimensional regularization ($D=4-2\varepsilon$), and 
before renormalization,
\begin{equation}\label{leff1lo}
{\cal L}_{eff,1} =-\frac{M^2}{16\pi^2}\left(\bar A+\bar B-
\frac{M^2}{2f^2}(\alpha^2+\beta^2-f^2)\right)
\left(\frac{1}{\varepsilon}-\gamma+\ln 4\pi +\ln\frac{\mu^2}{M^2}+1\right)
\end{equation}
\begin{equation}\label{leff2lo}
{\cal L}_{eff,2} =\frac{M^2}{32\pi^2 f^2}((2\bar B+m^2)\alpha^2 +
2\bar A \beta^2 - 4\bar C\alpha\beta)
\left(\frac{1}{\varepsilon}-\gamma+\ln 4\pi +\ln\frac{\mu^2}{M^2}+1\right)
\end{equation}
up to terms of order $v^4/16\pi^2$.
The coefficients $\bar A$, $\bar B$, $\bar C$, $\alpha$ and $\beta$
are functions of the (background) Higgs field $h$.
Recall that $M^2(\alpha^2+\beta^2-f^2)/(2f^2)={\cal O}(v^2)$.


\end{document}